\documentclass[aps,prd,twocolumn,superscriptaddress,nofootinbib]{revtex4-2}

\usepackage[colorlinks=true,citecolor=red,urlcolor=blue]{hyperref}
\usepackage{amssymb}
\usepackage{amsmath,color}

\usepackage{graphicx}
\usepackage{bbm}
\usepackage{verbatim}
\usepackage[normalem]{ulem}
\usepackage[caption=false]{subfig}
\usepackage[T1]{fontenc}
\usepackage[utf8]{inputenc}

\usepackage{url}
\usepackage{footnote}
\makesavenoteenv{tabular}
\usepackage{dcolumn}
\usepackage{bm}
\usepackage{xcolor}
\usepackage{multirow}
\usepackage{orcidlink} 

\begin{document}


\title{
\textbf{Impact of higher-order modes on eccentricity measurement in binary black hole gravitational waves} 
}%

\author{Honglue Tang\orcidlink{0009-0003-9453-2779}}
\author{Jinzhao Yang\orcidlink{0000-0002-4826-6014}}%
\author{Baoxiang Wang\orcidlink{0009-0003-3998-4609}}
\author{Tao Yang\orcidlink{0000-0002-2161-0495}}
 \email{Contact author: yangtao@whu.edu.cn}
\affiliation{School of Physics and Technology, Wuhan University, Wuhan 430072, China
}%

\date{\today}

\begin{abstract}
We investigate the systematic biases in measuring orbital eccentricity for binary black hole (BBH) mergers that arise when higher-order modes (HOMs) of gravitational waves are neglected in waveform modeling. Using Bayesian inference with the state-of-the-art eccentric, spin-aligned, higher-mode effective-one-body model \texttt{SEOBNRv5EHM}, we reanalyze six previously suggested eccentric gravitational-wave events—GW190521, GW190620, GW190701, GW191109, GW200129, and GW200208\_222617. Comparing results with its dominant-mode-only counterpart \texttt{SEOBNRv5E}, we find no statistically significant HOM-induced bias in eccentricity for any of these events, including GW190521, whose eccentricity has been debated in the literature. To explore representative parameter regimes vulnerable to HOM omission, we perform a broad zero-noise injection campaign varying detector-frame total mass, mass ratio, eccentricity, inclination, and network SNR. We find that significant systematic biases ($\Delta_e/\sigma > 1$) arise predominantly in systems with relatively high total mass ($M^{\rm det}\gtrsim120M_\odot$), highly asymmetric mass ratios ($q \gtrsim 4$), large inclinations ($\theta_\textrm{JN} \gtrsim 30^\circ$), and high SNRs ($\rho^N_\textrm{mf}\approx50$). Notably, for quasicircular BBHs with approximately $M^{\rm det}\gtrsim140M_\odot$, our results suggest that neglecting HOMs may lead to strong false-positive evidence for nonzero eccentricity. By contrast, for lower-mass systems ($M^{\rm det}\sim100 M_\odot$), HOM exclusion produces negligible eccentricity biases. Our results demonstrate that although current eccentric candidates are not impacted by HOM omission, future eccentricity measurements—particularly for relatively massive, asymmetric, or edge-on systems—require HOM-inclusive waveforms to avoid substantial systematic errors.
\end{abstract}

\maketitle


\section{\label{sec:intro}Introduction}

Since the first detection of gravitational waves (GWs) in 2015~\citep{LIGOScientific:2016aoc}, the LIGO-Virgo-KAGRA (LVK) Collaboration~\citep{LIGOScientific:2014pky,VIRGO:2014yos,Aso:2013eba} has released four Gravitational-Wave Transient Catalogs, reporting more than 200 GW events up to the first part of the fourth observing run (O4a)~\citep{LIGOScientific:2018mvr,LIGOScientific:2020ibl,LIGOScientific:2021usb,KAGRA:2021vkt,LIGOScientific:2025slb}. These observations have enabled many applications in cosmology, astrophysics, and fundamental physics, such as the constraint of cosmic expansion~\citep{Schutz:1986gp,LIGOScientific:2017adf,DES:2019ccw,LIGOScientific:2021aug,LIGOScientific:2025jau}, population studies of compact binaries~\citep{LIGOScientific:2018jsj,LIGOScientific:2020kqk,KAGRA:2021duu,LIGOScientific:2025pvj}, and test of general relativity (GR) and black hole physics~\citep{Will:2014kxa,LIGOScientific:2016lio,LIGOScientific:2018dkp,LIGOScientific:2019fpa,LIGOScientific:2020tif,LIGOScientific:2021sio,Krishnendu:2021fga,LIGOScientific:2025rid,theligoscientificcollaboration2025blackholespectroscopytests,Bhat:2024hyb}. Most GW events observed to date are attributed to the compact binary coalescences (CBCs), including binary black holes (BBHs), binary neutron stars and neutron star-black hole systems. However, their astrophysical formation channels remain uncertain.

There are two leading formation channels proposed for CBCs: isolated evolution and dynamical formation~\citep{Bethe:1998bn,Belczynski:2001uc, Rodriguez:2016vmx,Mandel:2009nx,KAGRA:2021duu,Stevenson:2017tfq,Belczynski:2014iua,Mapelli:2020vfa,Bouffanais:2021wcr}. In isolated evolution, two stars evolve together, form black holes, and finally coalesce with GW emission~\citep{1988ApJ...329..764L, Bethe:1998bn, Ivanova:2012vx, Belczynski:2014iua, Belczynski:2016obo,Marchant:2016wow,Stevenson:2017tfq}. Over long evolution timescales, the orbital eccentricity of binaries efficiently dissipates with GW emissions, leading to nearly circularized orbits by the time the GW frequency $f$ enters the observing band of ground-based GW detectors like LIGO (e.g., $f\gtrsim10~\textrm{Hz}$)~\citep{Peters:1964zz,Kowalska:2010qg}. Dynamical formation refers to coalescences formed via dynamical interactions~\citep{Sigurdsson:1993zrm,PortegiesZwart:1999nm,Zevin:2020gbd}, including dynamical capture~\citep{OLeary:2008myb,Kocsis:2011jy,Gondan:2017wzd} and multibody encounters~\citep{Samsing:2013kua,Samsing:2017rat,Michaely:2019aet}, which usually occur in dense environments such as active galactic nuclei (AGN) and globular clusters~\citep{OLeary:2005vqo,Mckernan:2017ssq,Yang:2019okq}, as dynamical interactions happen more frequently in such environments. Unlike isolated evolution, binaries formed dynamically can retain measurable orbital eccentricity in the observing band due to the short duration from encounter to merger~\citep{Samsing:2017xmd,Zevin:2018kzq,Gondan:2018khr} or external disturbance such as the Kozai-Lidov oscillations~\citep{Kozai:1962zz,Lidov:1962wjn,Antonini:2012ad,Antognini:2013lpa,VanLandingham:2016ccd,Antonini:2017ash}. Therefore, orbital eccentricity provides a useful tool to distinguish astrophysical formation channels~\citep{Zevin:2021rtf}. Beyond its importance in astrophysics, some studies have pointed out that eccentricity can greatly improve the parameter estimation (PE) and localization of GW sources~\citep{Sun:2015bva,Ma:2017bux,Pan:2019anf,Yang:2022tig,Yang:2022iwn,Yang:2022fgp,Zhong:2026pjg}. Moreover, \citet{Yang:2023zxk,Yang:2024vfy} proposed that eccentricity-induced higher harmonics can significantly enhance early warning of the detection and localization of GWs. On the other hand, ignoring eccentricity can induce systematic biases in PE and loss in signal recovery, affecting the GW application in astrophysics and test of GR~\citep{Favata:2013rwa,GilChoi:2022waq,Saini:2022igm,Narayan:2023vhm,Divyajyoti:2023rht,Gadre:2024ndy,Yang:2026mam,Bhat:2022amc,Saini:2023rto}. Meanwhile, the accurate identification of eccentricity itself is also challenging, since it has been pointed out that multiple effects may mimic eccentricity, such as microlensing of GWs, line-of-sight acceleration, massive graviton effect, and dipole radiation~\citep{Bhat:2025lri,Tiwari:2025fua}. In the presence of such eccentricity mimickers, parameter estimation using eccentric GW waveforms may yield spurious nonzero eccentricity. Therefore, the precise and accurate measurement of orbital eccentricity and relevant data analysis have become an essential aspect in GW astronomy.

The LVK Collaboration has not confirmed any eccentric event with waveform-independent analysis~\citep{LIGOScientific:2019dag,LIGOScientific:2023lpe}. Nevertheless, some research groups have independently reported several eccentric candidates based on waveform-dependent methods. Applying the eccentric waveform model \texttt{SEOBNRE} with likelihood reweighting, \citet{Romero-Shaw:2020thy,Romero-Shaw:2021ual} identified GW190521 and GW190620 as eccentric candidates. They later reported two additional candidates, GW191109 and GW200208\_222617, with the same approach~\citep{Romero-Shaw:2022xko}. In parallel, \citet{Gayathri:2020coq} reported $e\approx0.69$ for GW190521 through a numerical-relativity (NR) method. Furthermore, a machine-learning-based analysis using \texttt{SEOBNRv4EHM} reported three potential eccentric events: GW190701, GW200129, and GW200208\_222617~\citep{Gupte:2024jfe}. The most recent analysis with the phenomenological, eccentric model \texttt{IMRPhenomTEHM} reanalyzed 17 events, reporting strong support for eccentricity in GW200129 and GW200208\_222617, with weaker evidence in GW190701 and GW190929~\citep{Planas:2025jny}. 

Among these eccentric candidates, GW190521 has some notable source properties. It was the most massive BBH observed at the time, with a total mass of 153 $M_\odot$ in the source frame~\citep{LIGOScientific:2020ufj,LIGOScientific:2021usb}. The remnant black hole provides the first clear evidence for an intermediate-mass black hole, while the primary black hole is highly likely to lie within the pair-instability mass gap, suggesting a possible hierarchical merger. Moreover, \citet{Graham:2020gwr} reported a potential electromagnetic counterpart of GW190521 consistent with a kicked BBH merger in an AGN disk. All of the features mentioned above support the idea that GW190521 could have formed in a dynamical scenario, which in turn may suggest a nonzero eccentricity. Intriguingly, early analyses using \texttt{SEOBNRE} and NR support that GW190521 is eccentric~\citep{Romero-Shaw:2020thy,Gayathri:2020coq}. However, using later developed waveforms including higher-order modes (HOMs), which are the multipolar components of GWs beyond the dominant $(\ell,|m|)=(2,2)$ mode, subsequent studies found no sign of eccentricity in GW190521~\citep{Iglesias:2022xfc,Ramos-Buades:2023yhy,Gupte:2024jfe,Gamboa:2024hli}. This is a well-known tension of the measurement of eccentricity in current GW catalogs. Such a tension could result from the incompleteness of waveform models, differences in analysis methods, or other unknown effects. Some studies have shown that omitting HOMs in waveforms may lead to significant systematic biases in parameter estimation~\citep{Varma:2014jxa,Chandramouli:2024vhw,Yi:2025pxe}. We note that this tension originates from the conflicting results between Refs.~\citep{Romero-Shaw:2020thy} and~\citep{Iglesias:2022xfc,Ramos-Buades:2023yhy,Gupte:2024jfe,Gamboa:2024hli}. The former used (2,2)-mode-only waveform \texttt{SEOBNRE}, while the latter employed HOM-included waveform models, such as \texttt{SEOBNRv4EHM}, \texttt{TEOBResumSGeneral}, and \texttt{SEOBNRv5EHM}. Therefore, it is important to investigate whether neglecting HOMs in waveform models could account for this tension.

Eccentric waveform modeling has advanced considerably from early post-Newtonian (PN) inspiral-only templates~\citep{Yunes:2009yz,Cornish:2010cd,ShapiroKey:2010cnz,Huerta:2014eca,Loutrel:2017fgu,Tanay:2016zog,Moore:2016qxz,Moore:2018kvz,Moore:2019xkm,Klein:2018ybm,klein2021efpeefficientfullyprecessing} to current inspiral-merger-ringdown (IMR) models which are calibrated with NR~\citep{Huerta:2016rwp,Huerta:2017kez,Hinder:2017sxy,Cao:2017ndf,Liu:2019jpg,Liu:2021pkr,Ramos-Buades:2019uvh,Ramos-Buades:2021adz,Islam:2021mha,Nagar:2021gss,Nagar:2024dzj,Chattaraj:2022tay,Manna:2024ycx,Paul:2024ujx,Gamboa:2024hli,Planas:2025feq,Islam:2024zqo,Islam:2025rjl,Islam:2025llx,Islam:2025bhf}. A general approach for many recent eccentric IMR models employs the effective-one-body (EOB) formalism, which combines PN dynamics with NR calibration to construct semianalytical waveform models. Representative waveform families include \texttt{SEOBNR} and \texttt{TEOBResumS}. Early EOB models were based on the spin-aligned assumption, with spin precession and HOMs omitted. (e.g., \texttt{SEOBNRE}~\citep{Cao:2017ndf,Liu:2019jpg}). More current EOB eccentric models incorporate aligned (or misaligned) spins and HOMs (e.g., \texttt{SEOBNRv5EHM}~\citep{Gamboa:2024hli} and \texttt{TEOBResumS-Dali}~\citep{Nagar:2024dzj}), with a model developed from \texttt{SEOBNRE} further including spin precession~\citep{Liu:2023ldr}. In parallel, a phenomenological, spin-aligned, HOM-included, eccentric IMR model, \texttt{IMRPhenomTEHM}, has recently been proposed, providing an efficient alternative for eccentric GW studies~\citep{Planas:2025feq}.

These developments on waveforms make it possible to study eccentricity and HOMs within a common framework, which motivates us to investigate the important question of whether HOM omission can explain the tension of GW190521's eccentricity. Meanwhile, although several studies have assessed systematic biases in PE due to neglecting HOMs for other parameters, the specific impact of HOM omission on eccentricity measurement remains unclear. Therefore, we also intend to investigate whether HOM omission can lead to significant biases in the estimation of eccentricity, in which representative parameter regimes such biases may arise, and how they generally depend on source parameters. Finally, since Bayesian PE with current HOM-included eccentric waveforms remains computationally expensive, eccentric waveforms without HOMs may still be practical. Therefore, answering these questions can also provide practical guidance on when the use of faster eccentric waveforms without HOMs is appropriate, and when the use of HOM-included waveforms is necessary for robust and accurate eccentricity estimation.

For the computational efficiency and model accuracy, we employ the state-of-the-art eccentric waveform \texttt{SEOBNRv5EHM}~\citep{Gamboa:2024hli} to analyze the systematic biases in the measurement of orbital eccentricity due to omitting HOMs in the GW waveforms. We first investigate whether the tension in the measurement of GW190521's eccentricity is caused by omitting the HOMs. To do this, we employ \texttt{SEOBNRv5EHM} and the dominant-only modes of \texttt{SEOBNRv5EHM} ($(\ell,|m|)=(2,2)$, henceforth referred to as \texttt{SEOBNRv5E}) to reanalyze GW190521 and compare the constraints of eccentricity with the two models. Beyond GW190521, we also reanalyze the other five eccentric candidates and assess whether their eccentricity measurements are affected by the HOMs. Moreover, we explore the representative regimes where neglecting HOMs may introduce significant systematic biases in eccentricity, and qualitatively study the dependence of these biases on important parameters. In particular, we carry out a systematic injection-recovery study by broadly sampling over five parameters: the total mass $M=m_1+m_2$, mass ratio $q=m_1/m_2\ge1$, orbital eccentricity $e$, inclination angle $\theta_{\textrm{JN}}$, and matched-filter network SNR $\rho^N_\textrm{mf}$. The injections are based on the LIGO and Virgo detector network for consistency with GW detections, and the injected luminosity distance is adjusted to achieve the desired network SNR. For each injection, we generate HOM-included waveforms employing \texttt{SEOBNRv5EHM} and recover parameters using \texttt{SEOBNRv5E}.

This paper is organized as follows. Section~\ref{sec:method} describes our methodology. The analysis of eccentric GW candidates is presented in Sec.~\ref{RW}, followed by injection studies in Sec.~\ref{IA}. In Sec.~\ref{sec:cd}, we conclude the results with further discussions.

\section{\label{sec:method}Methodology}

\subsection{Eccentric waveform models}

We employ the time-domain EOB waveform model \texttt{SEOBNRv5EHM} implemented in the \texttt{pySEOBNR} package, which is designed for spin-aligned BBHs incorporating eccentricity and higher-order modes~\citep{Mihaylov:2023bkc,Gamboa:2024hli}. For parameter estimation, time-domain waveforms are converted to the frequency domain via the Fourier transform. As a member of the state-of-the-art \texttt{SEOBNR} family, \texttt{SEOBNRv5EHM} targets eccentric BBHs with moderate eccentricity and has been validated up to $e\approx0.5$ at $f=20~\textrm{Hz}$, being among the most accurate eccentric waveform models available at the time of publication.

\texttt{SEOBNRv5EHM} describes eccentric binaries using a Keplerian parametrization of the eccentric orbit:
\begin{equation}
    r = \frac{M}{u_p(1+e\cos{\zeta})}\,,
\end{equation}
where $M=m_1+m_2$ represents the total mass in geometric units with $m_1,m_2$ being the component masses, $e$ corresponds to the Keplerian orbital eccentricity, $\zeta$ represents the relativistic anomaly, and $u_p$ is the inverse semilatus rectum. All parameters are specified at the user-defined starting frequency $f_0$, which corresponds to the dimensionless orbit-averaged orbital frequency $\langle M\Omega\rangle$, and $\Omega$ represents the instantaneous orbital frequency. For simplicity, all eccentricities and relative anomalies discussed hereafter are defined at the starting frequency, denoted as $e_0$ and $\zeta_0$. 

\texttt{SEOBNRv5EHM} contains the dominant $(\ell,|m|) =$ (2,2) mode and HOMs $(\ell,|m|) =$ {(2,1),(3,3),(3,2),(4,4),(4,3)}. It constructs the gravitational polarizations according to this equation:
\begin{equation}
    h_+-ih_\times = \sum_{l=2}^\infty\sum_{m=-l}^{l}{}_{-2}Y_{\ell m}{(\theta_{\text{JN}},\varphi)}h_{lm}(\Theta,t)\,,
\end{equation} where $_{-2}Y_{\ell m}$ are the spherical harmonics of $-2$ spin weight depending on the inclination angle $\theta_\textrm{JN}$ and azimuthal angle $\varphi$, $h_{\ell m}$ are the gravitational-wave modes, and $\Theta$ represents the binary's intrinsic parameters. These modes include 3PN eccentricity corrections to the EOB radiation reaction force and the inspiral phase, while the merger-ringdown phase is modeled based on a quasicircular assumption. In the zero-eccentricity limit, \texttt{SEOBNRv5EHM} is demonstrated to be consistent with \texttt{SEOBNRv5HM}~\citep{Pompili:2023tna}. Comparing against NR eccentric waveforms from the Simulating eXtreme Spacetimes Collaboration, the (2,2) mode of \texttt{SEOBNRv5EHM} achieves a median unfaithfulness of $\sim0.02\%$ with a total mass range of $20-200~M_\odot$ and eccentricity up to 0.5 at $f=20~\textrm{Hz}$~\citep{Gamboa:2024hli}. Compared with its predecessor, \texttt{SEOBNRv4EHM}~\citep{Ramos-Buades:2021adz}, \texttt{SEOBNRv5EHM} improves in computational efficiency, model accuracy, and validated parameter space coverage. Under commonly used settings, \texttt{SEOBNRv5EHM} yields a slightly shorter PE runtime for GW150914 and GW190521 than \texttt{SEOBNRv4EHM\_opt}, an optimized version of \texttt{SEOBNRv4EHM}~\citep{Ramos-Buades:2023yhy,Gamboa:2024hli}. Based on our estimation, the improvement is approximately 2–5 times faster.

\subsection{Bayesian inference}
Bayesian inference is a fundamental tool for data analysis and parameter estimation in GW astronomy~\citep{Thrane:2018qnx,Romero-Shaw:2020owr}. According to Bayes' theorem, the probability distribution of parameter $\theta$ given measured GW signal $d$ and model $\mathcal{H}$ is obtained by 
\begin{equation}
    p(\theta|d,\mathcal{H}) = \frac{\mathcal{L}(d|\theta,\mathcal{H})\pi(\theta|\mathcal{H})}{\mathcal{Z}}\,,
\end{equation}
where $\pi(\theta|\mathcal{H})$ is the prior distribution of parameter $\theta$ given $\mathcal{H}$, $\mathcal{L}(d|\theta,\mathcal{H})$ is the likelihood function, and $\mathcal{Z}$ is the evidence of model $\mathcal{H}$ given measured data,
\begin{equation}
    \mathcal{Z} = \int{\mathcal{\mathcal{L}\left( \mathit{d} |\theta,\mathcal{H}\right )\pi\left(\theta|\mathcal{H}\right) \mathit{d}\theta}}\,.
\end{equation} We adopt a Gaussian likelihood function, assuming an N-detector network with zero-mean, stationary, Gaussian, and uncorrelated detector noise. The network log likelihood is defined as the sum of single-detector Gaussian log-likelihoods,
\begin{equation}
    \ln{\mathcal{L}(d|\theta,\mathcal{H})} \propto -\frac{1}{2}\sum_{i=1}^N\left\langle d - \mathcal{H}(\theta) \mid d - \mathcal{H}(\theta) \right\rangle_i\,,
\end{equation} where $i$ represents the $i$-th detector and N is the number of detectors. The noise-weighted inner product of two signals $A$ and $B$ is  defined as 
\begin{equation}
    \langle A \mid B \rangle_i \equiv4\Re\int_{f_\textrm{min}}^{f_{\textrm{max}}}{df\frac{\tilde{A}_i(f)\tilde{B_i^*}(f)}{S_n^{(i)}(f)}}\,,
\end{equation} where $S^{(i)}_n(f)$ is the one-sided noise power spectral density (PSD) of the $i$-th detector, and $\tilde{A}$ and $\tilde{B}$ represent the frequency-domain signal after the Fourier transform.

Parameter estimation is performed with \texttt{Bilby}~\citep{bilby_paper} and the nested sampler \texttt{dynesty}~\citep{Speagle:2019ivv}, using the following sampler settings: \texttt{sample=‘acceptance-walk’}, \texttt{naccept=60}, and \texttt{nlive=1000}. The sampling frequency is set to $f_\textrm{samp}=4096~\textrm{Hz}$. For GW event analysis, we set a minimum frequency $f_{\textrm{min}}= 20~\textrm{Hz}$ ($f_{\textrm{min}}= 11~\textrm{Hz}$ for GW190521, following~\citep{Romero-Shaw:2020thy,LIGOScientific:2020ufj,Gamboa:2024hli}). Meanwhile, the maximum frequency $f_{\textrm{max}}$ is 512, 448, 448, 448, 896, and 1792 $\textrm{Hz}$ for GW190521, GW190620, GW190701, GW191109, GW200129, and GW200208\_222617, respectively. For injection studies, we adopt $f_{\textrm{min}}=20~\textrm{Hz}$ and $f_{\textrm{max}}=2048~\textrm{Hz}$. 

The data segments are 8 s in duration, with a two-second postmerger duration, for all GW event analyses, except for GW190701 and GW191109,  for which four-second segments are used due to data quality issues. For injections, we adopt the same eight-second segments. When comparing HOM-included and HOM-excluded results for the same event or injection, we utilize identical sampler settings and analysis pipelines. Any unspecified sampler settings are set to the default values in \texttt{Bilby}. 

\subsection{Systematic bias}\label{systematic_bias}

\begin{figure*}[!htbp]
\includegraphics[width=0.9\textwidth]{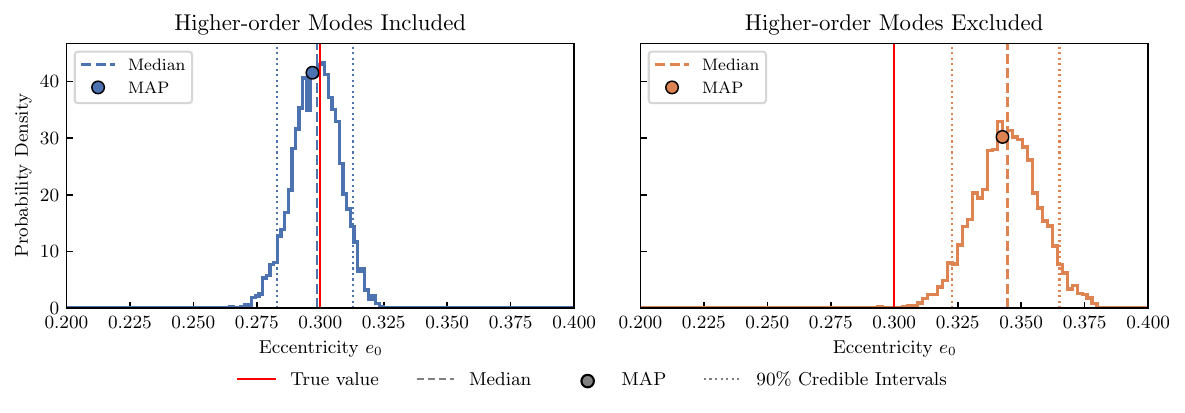}
\caption{The eccentricity posteriors for an example injected GW signal generated with \texttt{SEOBNRv5EHM}. The left panel shows the result recovered with \texttt{SEOBNRv5EHM}, while the right panel shows the result obtained with \texttt{SEOBNRv5E}. The red solid line indicates the true eccentricity. The colored dashed lines and circular markers show the median and MAP estimates of eccentricity, respectively. The colored dotted lines indicate the 90\% credible interval of eccentricity posteriors. As shown in the figures, the exclusion of HOMs significantly biases the estimated orbital eccentricity. This GW signal is generated with $M=160M_\odot$, $q=6$, $e_0=0.3$, $\theta_\textrm{JN}=90^\circ$, $\rho^N_\textrm{mf}=50$, and $f_\textrm{ref}=10~\textrm{Hz}$.\label{fig:bias}}
\end{figure*}
We quantify the impact of neglecting HOMs on eccentricity estimation primarily using the posterior-based systematic errors. Throughout this article, we use ``systematic bias'' to refer to the bias effect in PE caused by ignoring HOMs, and ``systematic error'' to describe the quantitative difference between the true values (or specified reference values) and predicted values.

Let the observed data $d$ be the combination of the true GW signal $s$ and detector noise $n$,
\begin{equation}
    d=s+n\,.
\end{equation} Assume $s$ can be explicitly described by a GW waveform model $h_\textrm{tr}$ given a set of true parameters $\boldsymbol{\lambda}_\textrm{tr}$ (i.e., $s=h_\textrm{tr}(\boldsymbol{\lambda}_\textrm{tr})$), and we use an approximate model $h_\textrm{app} (\boldsymbol{\lambda})$ to recover $\boldsymbol{\lambda}_\textrm{tr}$, resulting in the joint parameter posteriors $p_\textrm{app}(\boldsymbol{\lambda)}$ and estimated parameters $\boldsymbol{\lambda}_\textrm{app}$. When the approximate model exactly matches the true signal, the maximum \textit{a posteriori} (MAP) of $p_\textrm{app}(\boldsymbol{\lambda)}$ is expected to satisfy $\boldsymbol{\lambda}^\textrm{MAP}_\textrm{app}\approx\boldsymbol{\lambda}_\textrm{tr}$~\citep{Chandramouli:2024vhw,GilChoi:2022waq}. In contrast, the model difference between $h_\textrm{true}$ and $h_\textrm{app}$ can lead to biases in the estimation of $\boldsymbol{\lambda}_\textrm{tr}$. Figure~\ref{fig:bias} presents an example using HOM-excluded models to recover the orbital eccentricity of a HOM-included signal. We define the systematic error $\Delta_\lambda$ between the estimated parameters $\boldsymbol{\lambda}_\textrm{app}$ and true parameters $\boldsymbol{\lambda}_\textrm{tr}$ to quantify such biases:
\begin{equation}
    \Delta_\lambda=\boldsymbol{\lambda}_{\textrm{app}} - \boldsymbol{\lambda}_\textrm{tr}\,.
\end{equation}

In our injection analysis, the injected signals are generated using the HOM-included waveform model \texttt{SEOBNRv5EHM}, so the true parameters $\boldsymbol{\lambda}_\textrm{tr}$ are equal to the injected parameters $\boldsymbol{\lambda}_\textrm{inj}$ (i.e., $\boldsymbol{\lambda}_\textrm{tr}=\boldsymbol{\lambda}_\textrm{inj}$). We employ the HOM-excluded \texttt{SEOBNRv5E} as the approximate model for parameter recovery, leading to $\boldsymbol{\lambda}_{\textrm{app}}=\boldsymbol{\lambda}_{\textrm{22}}$, where $\boldsymbol{\lambda}_{\textrm{22}}$ is the parameters estimated with \texttt{SEOBNRv5E}. We report both the median-based and MAP-based systematic error
\begin{equation}
\Delta^\textrm{Median}_\lambda=\boldsymbol{\lambda}_{\textrm{22}}^{\textrm{Median}} - \boldsymbol{\lambda}_\textrm{inj}\,,~\Delta^\textrm{MAP}_\lambda=\boldsymbol{\lambda}^{\textrm{MAP}}_{\textrm{22}} - \boldsymbol{\lambda}_\textrm{inj}\,.
\end{equation}
$\boldsymbol{\lambda}^{\textrm{MAP}}$ is calculated as the MAP of the joint posterior, and $\boldsymbol{\lambda^\textrm{Median}}$ is the median values of corresponding one-dimensional marginalized posteriors. 

For GW events, the true value is actually unavailable. Therefore, we adopt \texttt{SEOBNRv5EHM} as an \textit{approximately true} model, with its corresponding estimated parameters $\boldsymbol{\lambda}_\textrm{HM}$ as a fiducial reference (i.e., $\boldsymbol{\lambda}_\textrm{tr}=\boldsymbol{\lambda}_\textrm{HM}$). Similarly, we compare HOM-included results $\boldsymbol{\lambda}_\textrm{HM}$ and HOM-excluded results $\boldsymbol{\lambda}_\textrm{22}$ using both MAP-based and median-based errors:
\begin{equation}
\Delta^\textrm{Median}_\lambda = \boldsymbol{\lambda_\textrm{22}^\textrm{Median}} - \boldsymbol{\lambda_\textrm{HM}^\textrm{Median}}\,,~
\Delta^\textrm{MAP}_\lambda = \boldsymbol{\lambda_\textrm{22}^\textrm{MAP}} - \boldsymbol{\lambda_\textrm{HM}^\textrm{MAP}}\,.
\end{equation}
In our study, we primarily rely on the median-based errors, with the MAP-based errors provided as a supplementary approach.

To evaluate the significance of systematic biases, we report the normalized systematic error $\Delta_\lambda/\sigma$, which is defined as the ratio of the systematic errors to their corresponding statistical errors $\sigma_\lambda$. Following~\citep{Chandramouli:2024vhw}, the statistical error $\sigma_{\lambda}$ of the parameter $\lambda$ is defined as
\begin{equation}
\sigma_{\lambda} = 
\begin{cases}
    Q_{100\%}(\lambda) - Q_{10\%}(\lambda), &\lambda_\textrm{app} > Q_{95\%}(\lambda) \\
    Q_{90\%}(\lambda) - Q_{0\%}(\lambda), &\lambda_\textrm{app} < Q_{5\%}(\lambda) \\
    \left[Q_{95\%}(\lambda) - Q_{5\%}(\lambda)\right]/2, & \text{else}
\end{cases}\,,
\end{equation} where $Q_{N\%}(\lambda)$ represents the $N\%$ percentile of $\lambda$ given its marginalized posterior $p(\lambda)$. Accordingly, the normalized systematic error of a specific parameter $\lambda$ (i.e., $e_0$) is derived by
\begin{equation}
    \Delta_{\lambda}/\sigma = \frac{|\Delta_{\lambda}|}{\sigma_{\lambda}}\,.
\end{equation} We assume that the systematic bias of ${\lambda}$ is significant when $\Delta_{\lambda}/\sigma>1$.

We also use the Jensen-Shannon divergence (JSD) as an auxiliary method to quantify the systematic bias for GW event analysis, which is a commonly used metric to quantify the difference (or similarity) of two distributions~\citep{Lin:1991zzm}. Consider two one-dimensional marginalized posteriors $p(\lambda)$ and $q(\lambda)$; the JSD between them is given by
\begin{equation}
    D_{JS}(p,q) = \frac{D_{KL}(p\|m)+D_{KL}(q\|m)}{2}\,,
\end{equation} where $m=(p+q)/2$, and $D_{KL}$ is the Kullback–Leibler divergence,
\begin{equation}
    D_{KL}(p\|q) = \int{d\lambda p(\lambda)\log\frac{p(\lambda)}{q(\lambda)}}\,.
\end{equation} The units of $D_{JS}$ and $D_{KL}$ correspond to the base of the logarithm (e.g., ``bits'' for the base of 2 and ``nats'' for the base of $e$). In this work, $D_{JS}$ is computed from one-dimensional marginalized posteriors of eccentricity and reported in bits, using a Gaussian kernel density estimator implemented in \texttt{SciPy}~\citep{2020SciPy-NMeth}. Following~\citep{Gupte:2024jfe,LIGOScientific:2020ibl,Romero-Shaw:2020owr}, we consider a JSD of 0.002 bits as the expected stochastic variation from the nested sampling, while a JSD larger than 0.007 bits suggests a significant bias between the compared distributions.

\section{Analysis of Eccentric GW Candidates}\label{RW}
\begin{figure*}[!t]
\includegraphics[width=0.9\textwidth ]{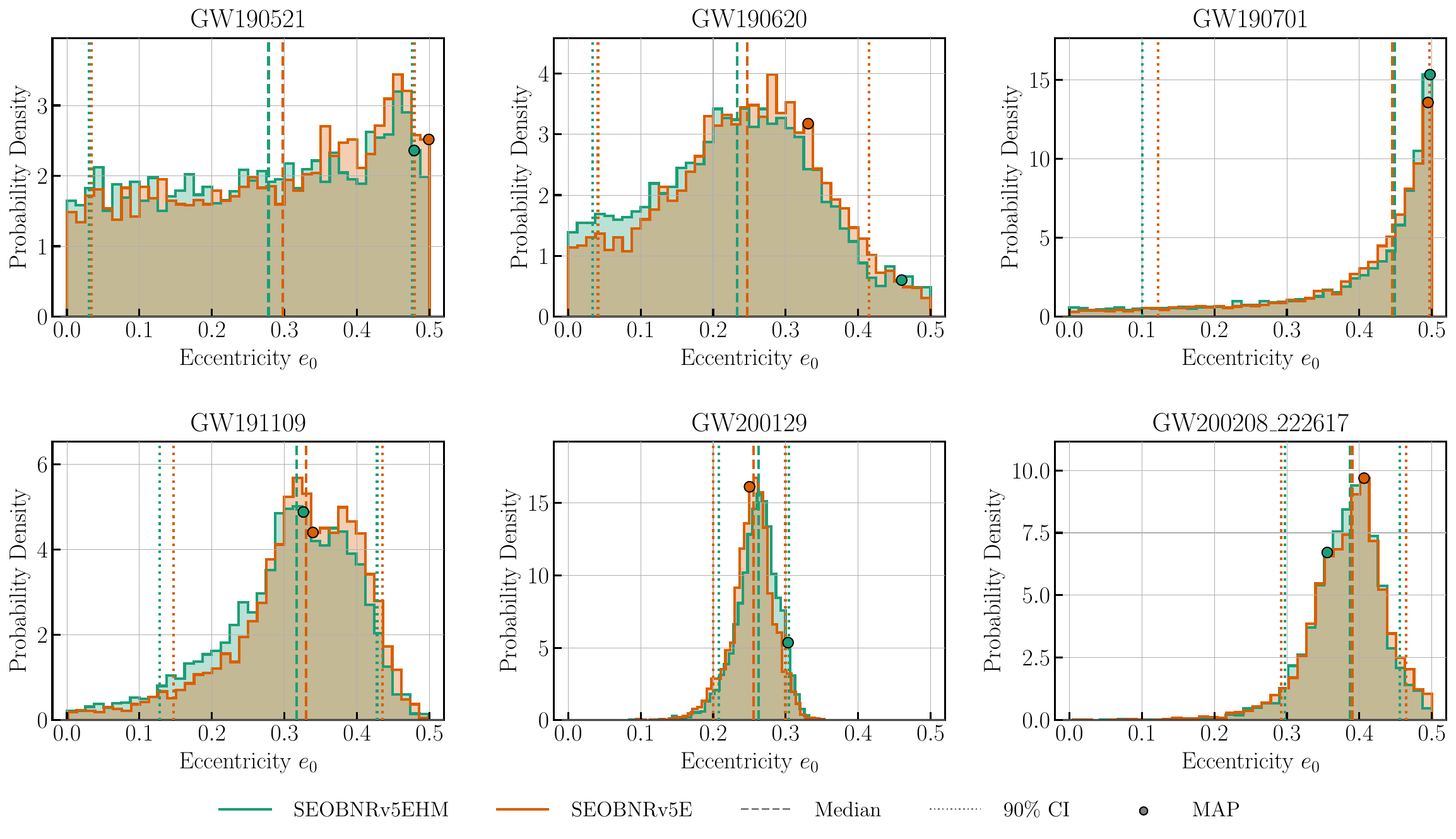}
\caption{Marginalized posterior distributions of initial eccentricity $e_0$ for all six GW events: GW190521, GW190620, GW190701, GW191109, GW200129, and GW200208\_222617. Each panel presents the eccentricity posteriors obtained with \texttt{SEOBNRv5EHM} (green) and \texttt{SEOBNRv5E} (orange). Colored dashed lines represent the median eccentricities estimated, and colored circular markers denote the MAP point for each model. Colored dotted lines indicate the 90\% credible interval of the eccentricity posteriors.\label{fig:real}}
\end{figure*}

\begin{table*}[!htbp]
\caption{Priors used for GW event analysis. \label{tab:real-parameters}
}
\begin{ruledtabular}
\renewcommand{\arraystretch}{1.2}
\begin{tabular}{cccc}
\textrm{Parameter}&
\textrm{Prior distribution}&
\textrm{Prior range}\\
\colrule
$\mathcal{M}$ $(M_\odot)$  & Uniform in component masses & Depends on different events \\
$1/q$  & Uniform in component masses & [0.05, 1] \\
$e_0$ & Uniform & [0.0, 0.5]  \\
$\zeta_0$ & Uniform & [0, $2\pi$]  \\
$d_L$ (Mpc) & $\propto d^2_L$\footnote{Uniform in comoving volume for GW190521.} & [$10^2$, $10^4$]  \\
$\theta_{\textrm{JN}}$ & Sine & [0, $\pi$]  \\
$\chi_1$ & AlignedSpin & [$-0.99$, $0.99$] \\
$\chi_2$& AlignedSpin & [$-0.99$, $0.99$] \\
$\phi$ & Uniform & [0, 2$\pi$]  \\
$\psi$ & Uniform & [0, $\pi$]  \\
R.A. & Uniform & [0, 2$\pi$] \\
Dec.  & Cosine & [$-\pi/2$, $\pi/2$] \\
$t_c$ (s) & Uniform & [$t^{\text{trigger}}-0.1$, $t^{\text{trigger}}+0.1$] \\
\end{tabular}
\end{ruledtabular}
\end{table*}

\begin{table*}[!htbp]
\caption{\label{tab:real}%
Eccentricity measurement for each model and event. $e^\textrm{Median}_0$ represents the median with 90\% credible interval (CI) and $e^{\rm MAP}_0$ is the maximum \textit{a posteriori} (MAP) estimate of eccentricity, both specified at $f_\textrm{ref}$; ${\Delta}_e/\sigma$ are the normalized systematic errors between \texttt{SEOBNRv5EHM} (HM) and \texttt{SEOBNRv5E} (22) based on both MAP and median; JSD denotes the Jensen-Shannon divergence between their eccentricity posteriors; and $\log_{10}\mathcal{B}_{\textrm{HM/22}}$ refers to the log-10 Bayes factors between the HOM-included (HM) and HOM-excluded (22) hypotheses.
}
\centering
\setlength{\tabcolsep}{10pt}
\begin{tabular}{c c c c c c c c}
\hline\hline
\\[-8pt]
Event &
Model &
$e^\textrm{Median}_0$ (90\% CI)&
$e^{\rm MAP}_0$ &
${\Delta}^\textrm{Median}_e/\sigma$ &
${\Delta}^\textrm{MAP}_e/\sigma$ &
JSD (bits) &
$\log_{10}\mathcal{B}_{\textrm{HM/22}}$ \\
\hline
\\[-6pt]
\multirow{2}{*}{GW190521} & HM & $0.28^{+0.20}_{-0.25}$ & 0.48 &
  \multirow{2}{*}{0.09} & \multirow{2}{*}{0.05} &
  \multirow{2}{*}{0.001} & \multirow{2}{*}{0.19} \\[4pt]
 & 22   & $0.30^{+0.18}_{-0.26}$ & 0.50 &  &  &  &  \\[3pt]
\hline
\\[-6pt]
\multirow{2}{*}{GW190620} & HM & $0.23^{+0.18}_{-0.20}$ & 0.46 &
  \multirow{2}{*}{0.07} & \multirow{2}{*}{0.69} &
  \multirow{2}{*}{0.003} & \multirow{2}{*}{-0.04} \\[4pt]
 & 22   & $0.25^{+0.17}_{-0.21}$ & 0.33 &  &  &  &  \\[3pt]
\hline
\\[-6pt]
\multirow{2}{*}{GW190701} & HM & $0.45^{+0.05}_{-0.35}$ & 0.50 &
  \multirow{2}{*}{0.02} & \multirow{2}{*}{0.01} &
  \multirow{2}{*}{0.002} & \multirow{2}{*}{-0.29} \\[4pt]
 & 22   & $0.45^{+0.05}_{-0.32}$ & 0.49 &  &  &  &  \\[3pt]
\hline
\\[-6pt]
\multirow{2}{*}{GW191109} & HM & $0.32^{+0.11}_{-0.19}$ & 0.33 &
  \multirow{2}{*}{0.09} & \multirow{2}{*}{0.09} &
  \multirow{2}{*}{0.006} & \multirow{2}{*}{0.41} \\[4pt]
 & 22   & $0.33^{+0.11}_{-0.18}$ & 0.34 &  &  &  &  \\[3pt]
\hline
\\[-6pt]
\multirow{2}{*}{GW200129} & HM & $0.26^{+0.04}_{-0.05}$ & 0.30 &
  \multirow{2}{*}{0.13} & \multirow{2}{*}{1.07} &
  \multirow{2}{*}{0.013} & \multirow{2}{*}{0.60} \\[4pt]
 & 22   & $0.26^{+0.04}_{-0.06}$ & 0.25 &  &  &  &  \\[3pt]
\hline
\\[-6pt]
\multirow{2}{*}{GW200208\_222617} & HM & $0.39^{+0.07}_{-0.09}$ & 0.36 &
  \multirow{2}{*}{0.03} & \multirow{2}{*}{0.59} &
  \multirow{2}{*}{0.002} & \multirow{2}{*}{0.20} \\[4pt]
 & 22   & $0.39^{+0.07}_{-0.10}$ & 0.41 &  &  &  &  \\[3pt]
\hline\hline
\end{tabular}
\end{table*}
\subsection{Event selection and parameter space}

We select six eccentric candidates: GW190521, GW190620, GW190701, GW191109, GW200129, and GW200208\_222617. Each has been reported to support the eccentric-BBH hypothesis in at least one study~\citep{Romero-Shaw:2020thy,Gayathri:2020coq,Romero-Shaw:2021ual,Romero-Shaw:2022xko,Gupte:2024jfe,Planas:2025jny}. These events span a broad range of parameter space: detector-frame (source-frame) total masses $M$ between 75 (63) $M_\odot$ and 243 (153) $M_\odot$, mass ratios $q$ from 1.17 to 4.66, initial eccentricities at adopted $f_\textrm{ref}$, $e_0$, from $\geq0.1$ up to 0.69, and network SNRs from 7.4 to 26.8. We use publicly available strain data from the Gravitational Wave Open Science Center (GWOSC) for parameter estimation, with PSDs estimated from the strain data using the Welch method~\citep{KAGRA:2023pio}. For GW events affected by glitches, we use the glitch-mitigated data products released by GWOSC~\citep{ligo_scientific_collaboration_and_virgo_2022_6477076,ligo_scientific_collaboration_and_virgo_2021_5546680}. Specifically, GW190701 and GW191109 use the \texttt{BayesWave} glitch modeling~\citep{Cornish:2014kda}, and GW200129 uses the \texttt{gwsubtract} linear subtraction~\citep{Davis:2018yrz,Davis:2022ird}. 

\subsection{Settings for analysis of eccentric candidates}

For all events except GW190521, parameters are specified at reference frequency $f_\textrm{ref}=10~\textrm{Hz}$, which is also the starting frequency for waveform generation. Following~\citep{LIGOScientific:2020ufj,Romero-Shaw:2020thy,Gamboa:2024hli}, for GW190521, $f_\textrm{ref}=5.5~\textrm{Hz}$ to ensure all waveform modes are in band at the likelihood minimum frequency $f_\textrm{min}=11~\textrm{Hz}$. As presented in Table~\ref{tab:real-parameters}, we adopt a detector-frame chirp mass prior $\mathcal{M^\textrm{det}}$ and an inverse mass ratio prior $1/q\in[0.05,1]$ that are implemented via uniform priors in component masses. For each event, the prior range of $\mathcal{M}$ is chosen based on the corresponding posterior samples released by GWOSC. In particular, following \citep{Gamboa:2024hli}, we set $\mathcal{M^\textrm{det}}\in[60,200]~M_\odot$ for GW190521. The priors for the initial eccentricity and relative anomaly at reference frequency are uniformly distributed with $e_0\in[0,0.5]$, $\zeta_0\in[0,2\pi]$. Following \citep{Gamboa:2024hli}, we set the luminosity distance prior to be proportional to $d^2_L$ with a range of $[10^2,10^4]$ Mpc for all events except GW190521. For GW190521, we instead adopt a prior uniform in comoving volume. Other priors are chosen as in~\citep{LIGOScientific:2018mvr,Gamboa:2024hli}. We enable distance and time marginalization to reduce computational cost.

Each event is analyzed twice, once with \texttt{SEOBNRv5EHM} and once with \texttt{SEOBNRv5E}. Subsequently, we compare the resulting posteriors from the two runs to calculate the normalized systematic errors ${\Delta}_e/\sigma$, the JSDs, and the Bayes factors $\log_{10}\mathcal{B}_{\textrm{HM/22}}$ between the HM and HOM-excluded (22) hypotheses.

\subsection{Results of eccentric candidates}

Figure~\ref{fig:real} presents the one-dimensional marginalized posteriors of $e_0$ obtained with \texttt{SEOBNRv5EHM} and \texttt{SEOBNRv5E}. Table~\ref{tab:real} lists the median eccentricities with 90\% credible interval and the MAP eccentricities, together with the corresponding normalized systematic errors, JSDs, and $\log_\textrm{10}$ Bayes factors. Across all six events, the measured values and posteriors of eccentricity are consistent with previous studies using \texttt{SEOBNRv4EHM} and \texttt{SEOBNRv5EHM}~\citep{Ramos-Buades:2023yhy,Gupte:2024jfe,Gamboa:2024hli}. Based on medians, we did not identify systematic biases in eccentricity attributable to the exclusion of HOMs. Notably, GW200129's ${\Delta}^\textrm{MAP}_e/\sigma$ ($1.07$) and JSD ($0.013$ bits) are beyond our thresholds for significant biases. However, its median-based normalized systematic error (${\Delta}^\textrm{Median}_e/\sigma=0.13$) is much smaller than the threshold. This event is noteworthy, but we do not consider the bias significant, as it does not show a measurable change in median eccentricities.

Event by event, neglecting HOMs leads to an increased median eccentricity for GW190521, GW190620, GW191109, and GW200208\_222617, and a decreased one for the rest of the events. GW190521, GW191109, and GW200208\_222617 exhibit an increased MAP eccentricity when HOMs are neglected, while the MAP eccentricity for the other three events decreases. GW200129 has the largest median-based normalized systematic error (${\Delta}^\textrm{Median}_e/\sigma = 0.13$), MAP-based systematic error (${\Delta}^\textrm{MAP}_e/\sigma=1.07$), and the largest JSD ($0.013$ bits). This may be due to its relatively high SNR ($26.8$) or the glitch in its data. In contrast, GW190701 exhibits the lowest median-based systematic error ${\Delta}^\textrm{Median}_e/\sigma=0.02$ and MAP-based systematic error (${\Delta}^\textrm{MAP}_e/\sigma=0.01$), which might be related to the railing posterior and boundary effect that greatly suppresses the true systematic error. GW190521 yields the smallest JSD ($0.001$ bits) and small normalized systematic errors, consistent with its uninformative and prior-driven eccentricity posterior. $\log_{10}\mathcal{B}_{\textrm{HM/22}}$ are close to zero for all events, also indicating that including or omitting HOMs has little impact on estimated parameters. Most events slightly prefer the HM hypothesis except for GW190620 and GW190701, which yield $\log_{10}\mathcal{B}_{\textrm{HM/22}}=-0.04$ and $-0.29$, respectively. Consistent with $\Delta_e/\sigma$ and JSD, GW200129 holds the largest $\log_{10}\mathcal{B}_{\textrm{HM/22}}$ of $0.60$.

In summary, the analysis of eccentric candidates reveals that none of the six events exhibits a significant systematic bias based on our joint criteria. In particular, for GW190521, we found no evidence of systematic biases in eccentricity estimation induced by omitting HOMs, suggesting that the exclusion of HOMs cannot explain the tension on its orbital eccentricity. In addition, to further validate the conclusion of the GW event analysis, we perform zero-noise injection studies based on the six events. The results support the conclusion drawn from our GW event analysis. Details of this verification are presented in the Appendix.

\section{Injection Analysis}\label{IA}
\subsection{Settings for injection analysis}

Although our analysis of the six GW events shows no clear evidence of significant systematic biases in eccentricity, we still need to determine under what conditions neglecting HOMs could significantly bias eccentricity measurements and, if they exist, how such biases depend on source parameters. To address this, we conduct an injection study over a grid of source parameters. Considering computational efficiency, the grid spans detector-frame total masses $M^\textrm{det}=\{100, 120, 140,160\}~M_\odot$ (Hereafter, unless otherwise specified, M refers to the detector-frame total mass), mass ratios $q={\text{\{1, 2, 4, 6\}}}$, initial eccentricities at reference frequency ($f_\textrm{ref}=10~\textrm{Hz}$) $e_0={\text{\{0, 0.3\}}}$, and inclination angles $\theta_{\text{JN}}=$\{$0^\circ$, $30^\circ$, $60^\circ$, $90^\circ$\}. We choose two network SNRs $\rho^N_{\text{mf}}={\text{\{20, 50\}}}$ by adjusting the injected luminosity distance. $\rho^N_{\text{mf}}$ is defined as
\begin{equation}
\rho^N_{\text{mf}}=\sqrt{(\rho^L_{\text{mf}})^2+(\rho^H_{\text{mf}})^2+(\rho^V_{\text{mf}})^2}\,,
\end{equation}
where $\rho^L_{\text{mf}}$, $\rho^H_{\text{mf}}$, and $\rho^V_{\text{mf}}$ represent the matched filter SNR of LIGO (Livingston), LIGO (Hanford), and Virgo, respectively. We use the Advanced LIGO and Virgo PSD at design sensitivity.  

These parameter choices are designed to broadly explore representative configurations where HOM effects are expected to be either weak or strong, allowing the exploration of how HOM-induced biases on eccentricity measurement change with different source properties. The effect of HOMs is more prominent for systems with highly asymmetric mass ratios and edge-on inclinations~\citep{Pekowsky:2012sr,Varma:2014jxa}. Also, several studies have identified significant systematic bias for parameter estimation when neglecting HOMs in the high-$M$ cases~\citep{Varma:2014jxa,Yi:2025pxe}. Hence, we systematically vary the injected $M$, $q$, and $\theta_\textrm{JN}$. In addition, because the influence of omitting HOMs may also depend on the true orbital eccentricity, we consider several injected eccentricities. The $e_0=0$ injections refer to quasicircular systems and enable us to broadly investigate whether omitting HOMs could lead to false claims of nonzero eccentricity for quasicircular systems. By contrast, the $e_0=0.3$ cases are chosen as a representative value for the eccentricities estimated for the six GW events analyzed above, allowing us to approximately quantify HOM-induced systematic biases for eccentric events. The reason for selecting different SNRs is that, although the absolute values of systematic errors are considered to be SNR independent~\citep{Chandramouli:2024vhw,Cutler:2007mi}, the normalized systematic errors can increase as the statistical errors are suppressed at high SNRs.

We employ \texttt{SEOBNRv5EHM} for injected signal generation and \texttt{SEOBNRv5E} for recovery, both with $f_\textrm{ref} = 10~\textrm{Hz}$, $f_\textrm{min}=20~\textrm{Hz}$, and $f_\textrm{max}=2048~\textrm{Hz}$ for consistency with our analysis of eccentric candidates and previous studies~\citep{Gamboa:2024hli}. To isolate systematic error from potential error introduced by detector noise, we perform zero-noise injections. In this paper, we focus on the systematic biases of eccentricity induced by HOMs. Thus, we fix component spins to zero to avoid the spin effect on the parameter estimation. 

We apply a detector-frame chirp mass prior $\mathcal{M}^\textrm{det}\in[5,100] ~M_\odot$ and an inverse mass ratio prior $1/q\in[0.05,1]$ that are uniformly sampled in component masses. The priors of the initial eccentricity and initial relative anomaly are uniformly distributed with a range of $e_0\in[0,0.5]$, $\zeta_0\in[0,2\pi]$. All the rest of the priors are identical to those adopted in our GW event analysis. We enable time, distance, and phase marginalization for computational efficiency. Table~\ref{tab:parameters} presents the injected values and prior for each parameter.

\begin{table*}[!htbp]
\caption{Injected values and priors of parameters for injection studies. \label{tab:parameters}
}
\begin{ruledtabular}
\renewcommand{\arraystretch}{1.2}
\begin{tabular}{ccccc}
\textrm{Parameter}&
\textrm{Injected values}&
\textrm{Prior distribution}&
\textrm{Prior range}\\
\colrule
$M$ $(M_\odot)$ & 100, 120, 140, 160 & $\cdots$ & $\cdots$ \\
$\mathcal{M}$ $(M_\odot)$ & $\cdots$ & Uniform in component masses & [5, 100]\\
$1/q$ & 1, 1/2, 1/4, 1/6 & Uniform in component masses & [0.05, 1]\\
$e_0$ & 0, 0.3 & Uniform & [0.0, 0.5] \\
$\zeta_0$ & 0 & Uniform & [0, 2$\pi$]  \\
$d_L$ (Mpc) & Depends on different events & $\propto d^2_L$ & [$10^2$, $10^4$] \\
$\theta_{\textrm{JN}}$ & $0^\circ$, $30^\circ$, $60^\circ$, $90^\circ$ & Sine & [0, $\pi$]  \\
$\phi$ & 0 & Uniform & [0, 2$\pi$] \\
$\psi$ & 0 & Uniform & [0, $\pi$] \\
R.A. & 2.2 & Uniform & [0, 2$\pi$] \\
Dec. & $-1.25$ & Cosine & [$-\pi/2$, $\pi/2$] \\
$t_c$ (s) & 1126259642.413 & Uniform & [$1126259642.313$, $1126259642.513$] \\
\end{tabular}
\end{ruledtabular}
\end{table*}


\begin{figure*}[!htbp]
\includegraphics[width=0.8\textwidth]{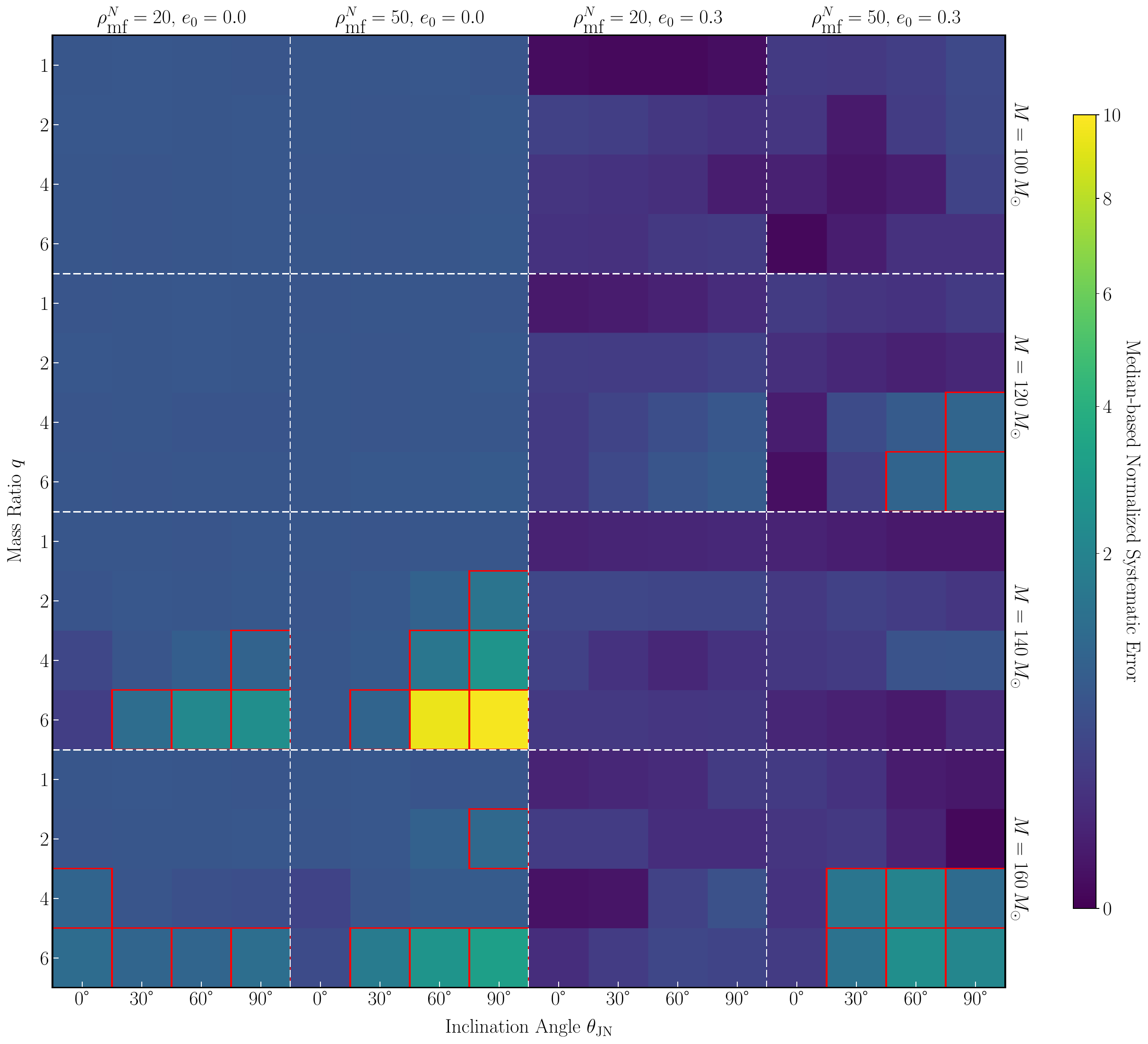}
\caption{Median-based normalized systematic errors $\Delta_e/\sigma$ for all injection analyses. Rows compare results with different total masses, and columns compare results with different network SNRs and eccentricities. The block colors correspond to the values of normalized $\Delta_e/\sigma$. Every block with $\Delta_e/\sigma>1$ is marked with a red box.\label{fig:norm-erro-median}}
\end{figure*}

\begin{figure*}[!htbp]
\includegraphics[width=0.8\textwidth]{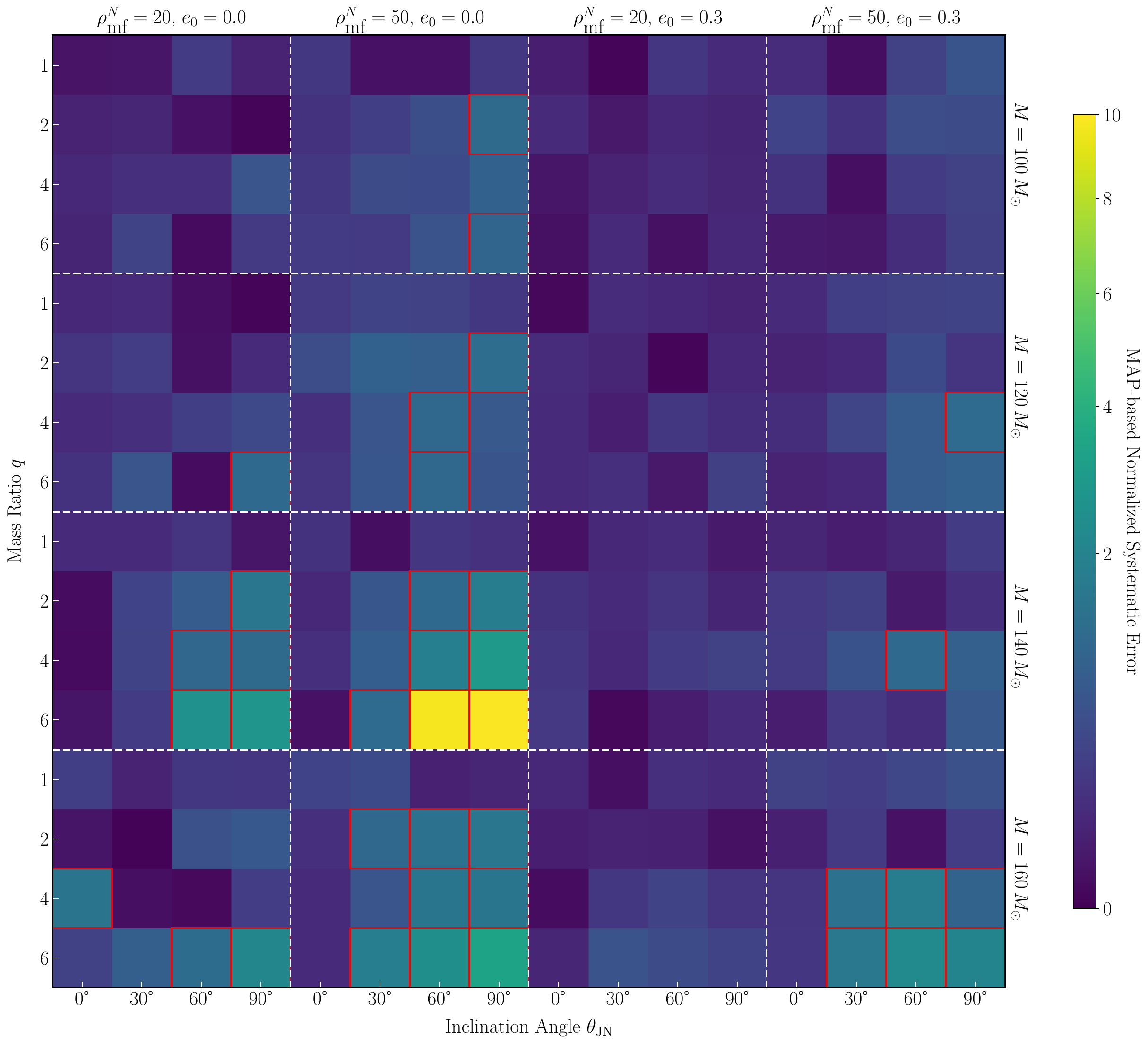}
\caption{The same as in Fig.~\ref{fig:norm-erro-median} but for normalized systematic errors calculated using MAP.\label{fig:norm-erro-map}}
\end{figure*}

\begin{figure*}[htbp]
\includegraphics[width=0.75\textwidth]{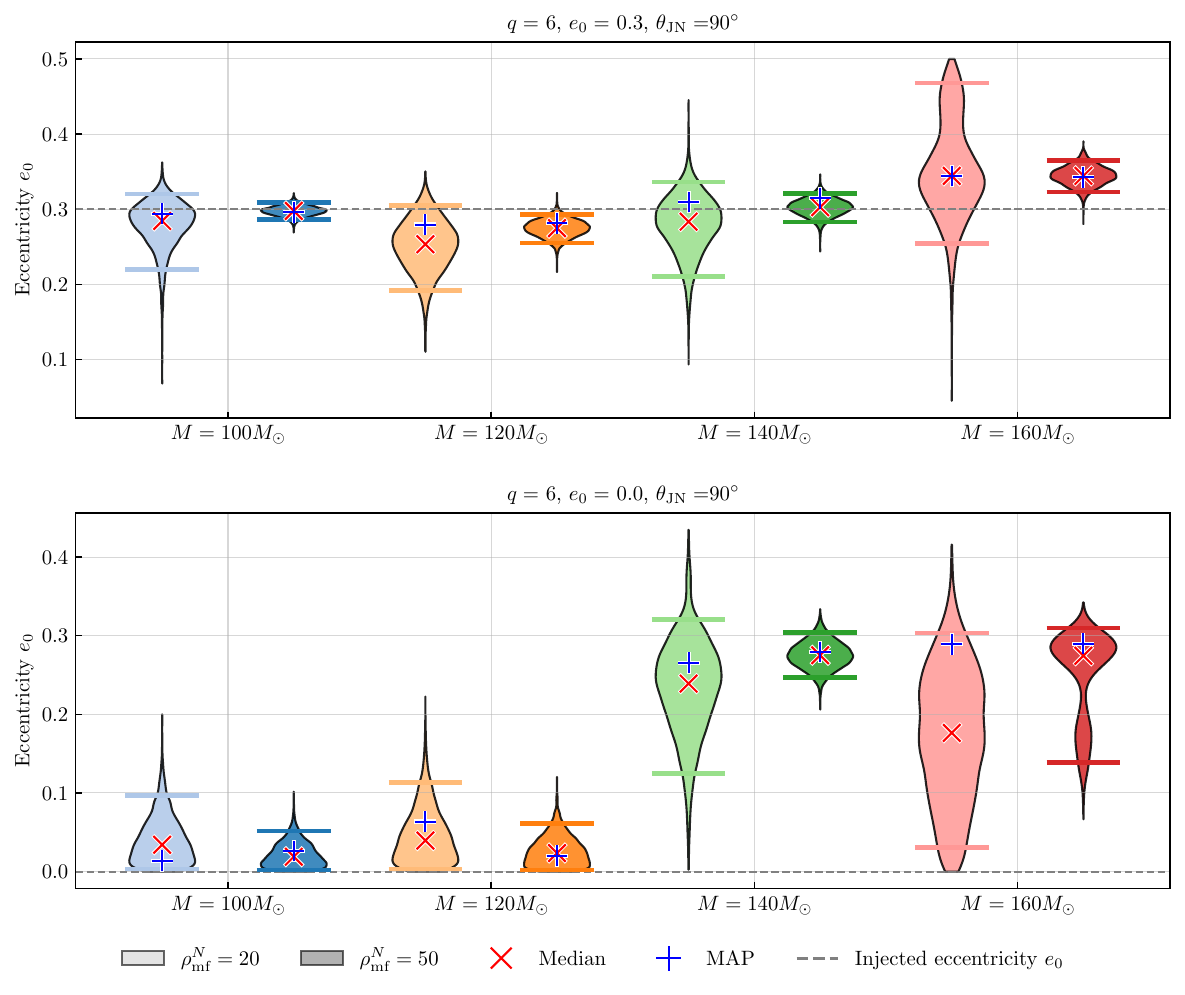}
\caption{Eccentricity posteriors recovered with HOM-excluded \texttt{SEOBNRv5E} for injections generated with HOM-included \texttt{SEOBNRv5EHM}. We show results for systems with different total masses $M$ and compare two network matched-filter SNRs, $\rho^N_\textrm{mf}=20$ (lighter color) and $\rho^N_\textrm{mf}=50$ (darker color). The top panel presents the case of $q=6$, $e_0=0.3$, and $\theta_\textrm{JN}=90^\circ$; the bottom panel presents the case of $q=6$, $e_0=0.0$, and $\theta_\textrm{JN}=90^\circ$. The gray dashed lines indicate injected values of eccentricity, and the colored solid lines represent the 90\% credible interval of each corresponding eccentricity posterior. The median and MAP eccentricities are marked with red crosses and blue plus signs, respectively.\label{Violins-M}}
\end{figure*}

\begin{figure*}[htbp]
\includegraphics[width=0.75\textwidth]{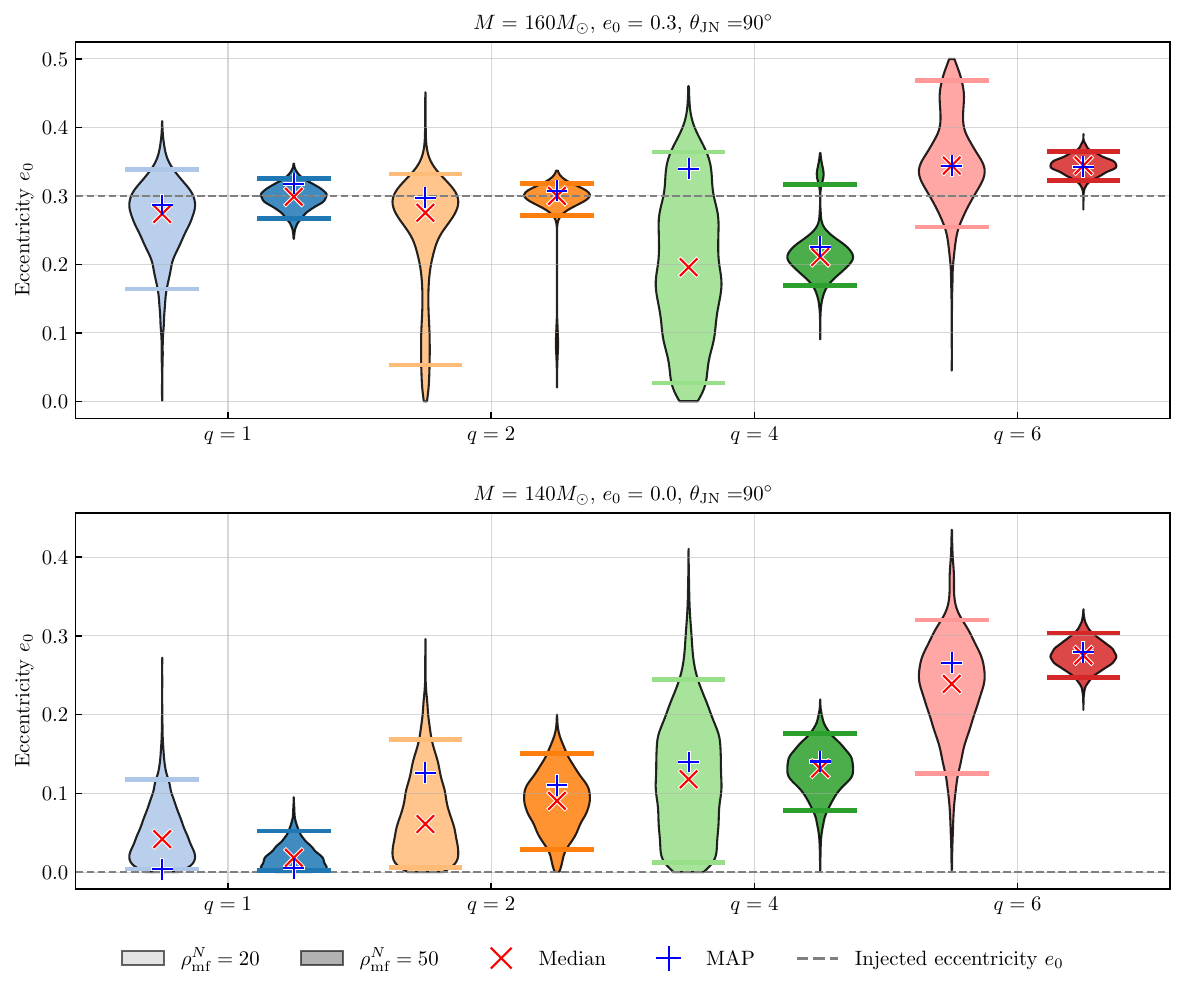}
\caption{The same as in Fig.~\ref{Violins-M} but for systems with different mass ratios $q$. The top panel presents the case of $M=160M_\odot$, $e_0=0.3$, and $\theta_\textrm{JN}=90^\circ$; the bottom panel presents the case of $M=140M_\odot$, $e_0=0.0$, and $\theta_\textrm{JN}=90^\circ$.\label{Violins-q}}
\end{figure*}

\begin{figure*}[htbp]
\includegraphics[width=0.75\textwidth]{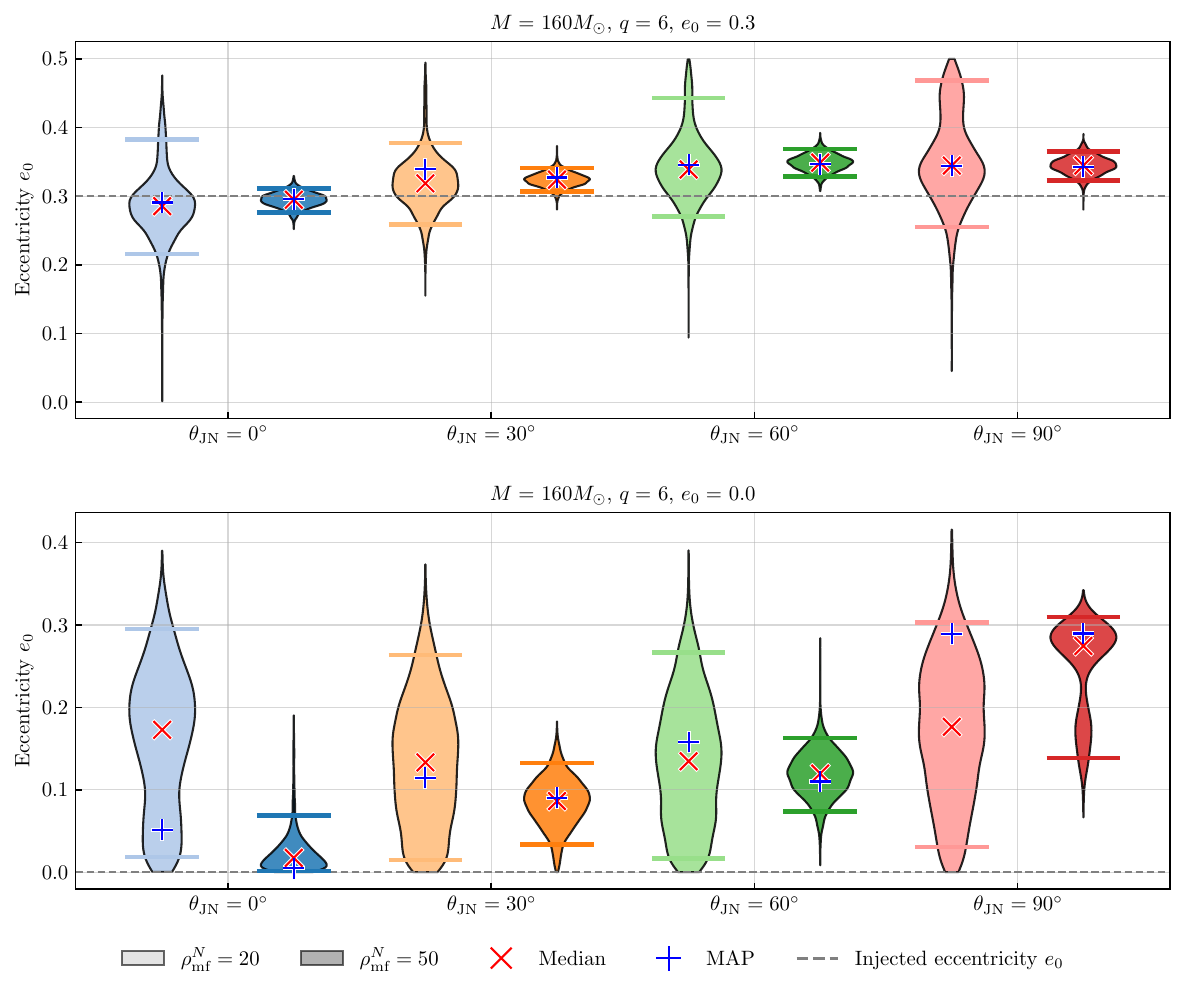}
\caption{The same as in Fig.~\ref{Violins-M} but for systems with different inclination angles $\theta_\textrm{JN}$. The top panel presents the case of $M=160M_\odot$, $q=6$, and $e_0=0.3$; the bottom panel presents the case of $M=160M_\odot$, $q=6$, and $e_0=0.0$.\label{Violins-iota}}
\end{figure*}
\subsection{Results for injection analysis}

Figures~\ref{fig:norm-erro-median} and~\ref{fig:norm-erro-map} show the normalized systematic error ${\Delta}_e/\sigma$ when HOMs are neglected in the GW waveforms. Red boxes mark the systems with ${\Delta}_e/\sigma>1$. These results reveal that neglecting HOMs can raise non-negligible systematic biases in eccentricity. Based on medians, the biases are generally pronounced for relatively high total masses ($M\gtrsim120~M_\odot$), high mass ratios ($q\gtrsim4$), large inclinations ($\theta_{\text{JN}}\gtrsim30^\circ$), and high network SNRs ($\rho^N_{\text{mf}}\approx20$ (for eccentric cases $\rho^N_{\text{mf}}\approx50$)). This phenomenon generally vanishes for $M=100M_\odot$. Notably, red boxes frequently appear at $e_0=0$ for $M=\{140,~160\}~M_\odot$. This suggests that, when analyzing quasicircular systems with high total masses, high mass ratios, and large inclination angles, neglecting HOMs may lead to false-positive claims of eccentricity.

We identify several patterns from these results.

\subsubsection{Relation between HOM-induced biases and total mass}

Previous studies have reported significant HOM-induced systematic biases in parameter estimation for large $M$ systems~\citep{Varma:2014jxa,Yi:2025pxe}, and our results are mostly consistent with such an opinion. As shown in Fig.~\ref{Violins-M}, the normalized systematic error ${\Delta}_e/\sigma$ for the $M=100M_\odot$ and $M=120M_\odot$ cases is generally lower than that for $M=140M_\odot$ and $M=160M_\odot$. However, the absolute systematic error presents an oscillatory pattern, with decreased systematic biases in particular conditions (e.g., $e=0.3,~M=140M_\odot$ and $e=0.0,~M=160M_\odot$). One potential explanation for this phenomenon is that the systematic bias in eccentricity is related to the difference between injection and recovery waveforms. As HOMs become more prominent at larger total masses, the oscillation in phase would lead to an oscillatory behavior in waveform differences and hence a nonmonotonic dependence of HOM-induced eccentricity bias on total mass. Other studies have also reported similar oscillatory patterns in systematic biases~\citep{Cutler:2007mi,DeLuca:2025bph,Yang:2026mam}.

\subsubsection{Relation between HOM-induced biases and mass asymmetry}

As presented in Figs.~\ref{fig:norm-erro-median}–\ref{fig:norm-erro-map} and \ref{Violins-q}, there is a generally monotonic relation between mass ratios and HOM-induced eccentricity biases. For most systems with non-negligible systematic biases in eccentricity, larger mass ratios generally correspond to larger normalized systematic errors. This is consistent with our expectation, as HOMs become more prominent at higher mass ratios. However, there are certain circumstances in which the dependence of the eccentricity bias on mass ratios exhibits a similar oscillatory pattern as mentioned above (e.g., $M=160M_\odot$), as shown in the upper plot of Fig.~\ref{Violins-q}.

\subsubsection{Relation between HOM-induced biases and inclination angle}

Roughly, ${\Delta}_e/\sigma$ increases with $\theta_{\text{JN}}$ for fixed $(M,~q,~\rho^N_{\text{mf}})$. Figure~\ref{Violins-iota} compares a set of four eccentricity posteriors with the same injected parameters but different inclination angles, providing a clear view of how the biases vary with inclination angles. This naturally aligns with our expectation, as the HOMs are enhanced with larger inclination angles.

\subsubsection{Relation between HOM-induced biases and orbital eccentricity}

For systems with fixed $(M,~\rho^N_{\text{mf}})$, significant biases in eccentricity occur more frequently for quasicircular cases than for eccentric ones. Furthermore, the largest normalized systematic error in eccentricity is observed at $e_0=0.0$ ($\Delta^\textrm{Median}_e/\sigma\approx9.7$ at $(M,~q,~\theta_\textrm{JN},~ \rho^N_{\text{mf}})=(140M_\odot,~6,~90^\circ,~50)$). This implies that, compared to eccentric systems, ignoring HOMs is more likely to yield spurious nonzero eccentricity in parameter estimation for quasicircular binaries. This may partly result from the prior-boundary effect, as the true eccentricity $e_0=0.0$ falls on the lower edge of the eccentricity prior.

\subsubsection{Relation between HOM-induced biases and network SNR}

We note that the HOM-induced systematic biases in eccentricity are greater at higher network SNR. Figures~\ref{Violins-M}–\ref{Violins-iota} show how increasing network SNRs amplify these biases. The absolute systematic errors change little with SNR, which is consistent with \citep{Chandramouli:2024vhw} and our expectation. Instead, higher SNR primarily reduces the statistical errors and, consequently, increases the normalized systematic errors in eccentricity estimation. 

Finally, we find that for $M=100M_\odot$ systems, neglecting HOMs generally would not lead to significant systematic biases across the grid. This suggests that the dominant $(2,2)$ mode alone might be sufficient to constrain $e_0$ for low-mass systems of the sources specified in this study. This is consistent with our expectations, since parameter estimation for low-mass binaries is dominated by the long inspiral phase, where HOMs contribute much less than during the merger ringdown.

For validation, we select two systems that exhibit large eccentricity biases when HOMs are neglected and perform additional injection-recovery studies using \texttt{SEOBNRv5EHM} for both injection and recovery. In particular, we choose $(M,~q,~e_0,~\theta_\textrm{JN},~\rho^N_\textrm{mf}) =(140M_\odot,~6,~0.0,~90^\circ,~50)$ and $(160M_\odot,~6,~0.3,~90^\circ,~50)$, for which $\Delta^\textrm{Median}_e/\sigma\approx9.7$ and 2.1, respectively, when HOMs are omitted. In both cases, the injected parameter values are accurately recovered, validating our analysis pipeline. The left plot of Fig.~\ref{fig:bias} presents the recovered eccentricity posterior at $(M,~q,~e_0,~\theta_\textrm{JN},~\rho^N_\textrm{mf}) =(160M_\odot,~6,~0.3,~90^\circ,~50)$ when HOMs are included in PE.

\section{Conclusion and Discussion}\label{sec:cd}

In this work, we investigate the systematic biases in the measurement of the orbital eccentricity of binary black holes that are introduced by neglecting higher-order modes in the gravitational-wave waveforms. We perform Bayesian inference with \texttt{Bilby} and the spin-aligned eccentric IMR waveform model \texttt{SEOBNRv5EHM}. We quantify the biases using two metrics, including the MAP/median-based normalized systematic errors $\Delta_e/\sigma$ and the Jensen-Shannon divergence in bits.

In Sec.~\ref{RW}, we analyze six GW events using \texttt{SEOBNRv5EHM} and its HOM-excluded counterpart \texttt{SEOBNRv5E}, and compare the eccentricity posteriors obtained with the two models. We find no evidence of significant HOM-induced systematic biases ($\Delta_e/\sigma>1$) in eccentricity. Although GW200129 exhibits a MAP-based normalized systematic error $\Delta^\textrm{MAP}_e/\sigma$ of 1.07 and a JSD of 0.013 bits, its median eccentricity is mostly not affected by ignoring HOMs, with $\Delta^\textrm{Median}_e/\sigma=0.13$. Notably, the HOM-induced eccentricity bias of GW190521 is very small, suggesting that the ignorance of HOMs is insufficient to explain the tension of eccentricity measurement for GW190521. The discrepancy between different studies might be due to the choice of priors, waveform model errors, parameter estimation methods, or other unknown effects.

In Sec.~\ref{IA}, we present zero-noise injection studies to search for representative parameter regimes where HOM-induced biases in eccentricity are non-negligible ($\Delta_e/\sigma>1$), as well as to qualitatively investigate the trend of such biases on source parameters. We inject simulated GW signals using \texttt{SEOBNRv5EHM} and recover parameters using \texttt{SEOBNRv5E}, varying total mass, mass ratio, eccentricity, inclination angle, and network SNR. According to the results, neglecting HOMs could raise significant biases in eccentricity, which are prominent for systems generally characterized by relatively large total masses ($M\gtrsim120 M_\odot$), highly asymmetric mass ratios ($q\gtrsim4$), large inclinations ($\theta_{\text{JN}}\gtrsim30^\circ$), and high SNR ($\rho^N_{\text{mf}}\approx50$). Notably, quasicircular BBHs in this regime may yield false-positive claims of eccentricity when analyzed with HOM-excluding models. Therefore, to avoid systematic biases and false positives in eccentricity measurement, HOMs should be included in a future data analysis.

There are several limitations to our study. First, owing to waveform availability, we only compared HOM-included and HOM-excluded spin-aligned eccentric waveform models, without taking spin precession into account. As pointed out in \citet{Romero-Shaw:2020thy}, there is a degeneracy between spin-precession and eccentricity. Second, owing to the computational cost of \texttt{SEOBNRv5EHM}, we employ a parameter grid with limited resolution and coverage. A finer and broader grid search over these parameters would be necessary to draw a more comprehensive conclusion. With the development of faster and more accurate eccentric, HOM-included waveforms, such as the recently proposed \texttt{SEOBNRv6EHM}~\citep{Gamboa:2026jht}, this will become more feasible. We leave such an extension to future studies. Lastly, we adopt only one waveform model, which is based on the EOB formalism. It is worthwhile to compare HOM-induced eccentricity biases between different eccentric models (e.g., the phenomenological eccentric model \texttt{IMRPhenomTEHM}~\citep{Planas:2025feq}) for the robustness of the conclusion. In addition, the HOM-induced biases for third-generation ground-based GW detectors such as ET and CE, as well as space-based detectors LISA, Taiji, and TianQin, deserve in-depth study, as they are expected to operate with much higher sensitivity and a lower observing band than current detectors. We also leave this for future research.

\begin{acknowledgments}
We would like to thank Zhoujian Cao, Aldo Gamboa, Alexander Nitz, Shichao Wu, Keisi Kacanja, Kanchan Soni, Divyajyoti, Gareth Cabourn Davies, Tom Dent, and Yuxuan Li for their helpful discussions. We also thank Colm Talbot for assistance with \texttt{Bilby}.

This work is supported by the National Natural Science Foundation of China Grants No. 12575063, and in part by ``the Special Funds for the Double First-Class Development of Wuhan University'' under the Reference No. 2025-1302-010.

The numerical calculations in this paper have been done on the supercomputing system in the Supercomputing Center of Wuhan University. Some of the results in this paper have been derived using the ``\texttt{PESummary}'' package~\cite{pesummary}. This research has made use of data or software obtained from the Gravitational Wave Open Science Center~\cite{GWOSC}, a service of the LIGO Scientific Collaboration, the Virgo Collaboration, and KAGRA.
\end{acknowledgments}

\section*{Data Availability}
The data that support the findings of this article are not publicly available. The data are available from the authors upon reasonable request.

\appendix*
\section{Verification of analysis of eccentric candidates}\label{VerifRA}

\begin{figure*}[!hbtp]
\includegraphics[width=0.9\textwidth]{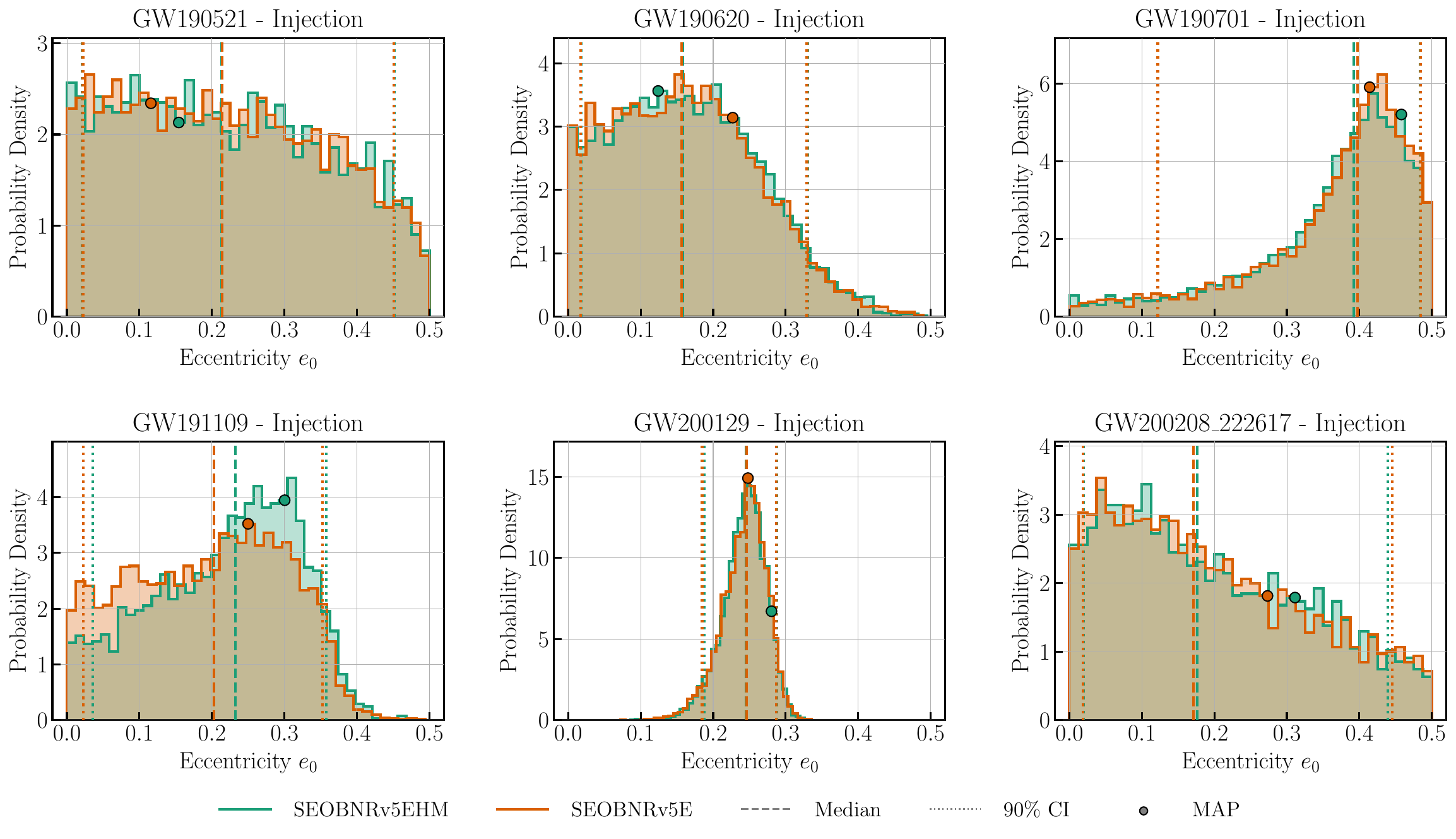}
\caption{The same as in Fig.~\ref{fig:real} but for event-based injection studies. \label{fig:event-based-injection}}
\end{figure*}

\begin{table}[htbp]
\centering
\begin{ruledtabular}
\begin{tabular}{ccccccc}
Event & 190521 & 190620 & 190701 & 191109 & 200129 & 200208 \\ \hline
\\[-6pt]
$\Delta_e^\textrm{Median}/\sigma$ & 0.00 & 0.01 & 0.03 & 0.18 & 0.02 & 0.02\\[2pt] 
$\Delta_e^\textrm{MAP}/\sigma$ & 0.18 & 0.66 & 0.24 & 0.31 & 0.63 & 0.18\\[2pt] 
JSD (bits) & 0.000 & 0.001 & 0.001 & 0.009 & 0.001 & 0.001\\[2pt]
$\log_{10}\mathcal{B}_{\textrm{HM/22}}$ & $-0.55$ & $-0.09$ & $-0.12$ & $-0.22$ & $-0.22$ & $-0.08$
\\[2pt]
\end{tabular}
\end{ruledtabular}
\caption{The results of event-based injections, including the normalized systematic errors in eccentricity based on the median and maximum \textit{a posteriori} (MAP) estimates, the Jensen-Shannon divergence (JSD), and the log-10 Bayes factors between the HM and HOM-excluded (22) hypotheses.\label{tab:event-based-injection}}
\end{table}

We perform event-based injection studies to verify the conclusion of the GW event analysis. First, we use \texttt{SEOBNRv5EHM} to generate simulated GW signals with median parameters of each event obtained in the GW event analysis using \texttt{SEOBNRv5EHM}. Next, we inject the signals into zero-noise detectors while using the same PSDs as those in the GW event analysis, and recover parameters with the identical settings of the GW event analysis using both \texttt{SEOBNRv5EHM} and \texttt{SEOBNRv5E}. Finally, we compare the results of the injection study and the GW event analysis to verify our conclusion. 

The results are shown in Table~\ref{tab:event-based-injection} and Fig.~\ref{fig:event-based-injection}. The event-based injections yield conclusions that omitting HOMs does not lead to significant biases in eccentricity measurement, which is consistent with our GW event analysis. Notably, GW191109-like injection exhibits a JSD of 0.009 bits that is beyond its threshold for significant biases. Nevertheless, the measured eccentricity, both for the median and MAP, of this injection is not significantly biased.

Notably, the eccentricity posteriors of GW190521, GW190701, and GW200208\_222617 in event-based injections are significantly different from those of the GW event analysis. This is not beyond our expectations due to the uninformative eccentricity posterior for GW190521, the railing posterior for GW190701, and the relatively low network SNR for GW200208\_222617. Moreover, the GW190521-like injection exhibits a badly converged eccentricity posterior. This also suggests an important implication: The orbital eccentricity of GW190521-like events may not be accurately constrained with the current sensitivity of LIGO and Virgo.

\bibliography{currentVer}

@article{LIGOScientific:2016aoc,
    author = "Abbott, B. P. and others",
    collaboration = "LIGO Scientific, Virgo",
    title = "{Observation of Gravitational Waves from a Binary Black Hole Merger}",
    eprint = "1602.03837",
    archivePrefix = "arXiv",
    primaryClass = "gr-qc",
    reportNumber = "LIGO-P150914",
    doi = "10.1103/PhysRevLett.116.061102",
    journal = "Phys. Rev. Lett.",
    volume = "116",
    number = "6",
    pages = "061102",
    year = "2016"
}

@article{LIGOScientific:2014pky,
    author = "Aasi, J. and others",
    collaboration = "LIGO Scientific",
    title = "{Advanced LIGO}",
    eprint = "1411.4547",
    archivePrefix = "arXiv",
    primaryClass = "gr-qc",
    doi = "10.1088/0264-9381/32/7/074001",
    journal = "Class. Quant. Grav.",
    volume = "32",
    pages = "074001",
    year = "2015"
}

@article{VIRGO:2014yos,
    author = "Acernese, F. and others",
    collaboration = "VIRGO",
    title = "{Advanced Virgo: a second-generation interferometric gravitational wave detector}",
    eprint = "1408.3978",
    archivePrefix = "arXiv",
    primaryClass = "gr-qc",
    doi = "10.1088/0264-9381/32/2/024001",
    journal = "Class. Quant. Grav.",
    volume = "32",
    number = "2",
    pages = "024001",
    year = "2015"
}

@article{Aso:2013eba,
    author = "Aso, Yoichi and Michimura, Yuta and Somiya, Kentaro and Ando, Masaki and Miyakawa, Osamu and Sekiguchi, Takanori and Tatsumi, Daisuke and Yamamoto, Hiroaki",
    collaboration = "KAGRA",
    title = "{Interferometer design of the KAGRA gravitational wave detector}",
    eprint = "1306.6747",
    archivePrefix = "arXiv",
    primaryClass = "gr-qc",
    doi = "10.1103/PhysRevD.88.043007",
    journal = "Phys. Rev. D",
    volume = "88",
    number = "4",
    pages = "043007",
    year = "2013"
}

@article{LIGOScientific:2018mvr,
    author = "Abbott, B. P. and others",
    collaboration = "LIGO Scientific, Virgo",
    title = "{GWTC-1: A Gravitational-Wave Transient Catalog of Compact Binary Mergers Observed by LIGO and Virgo during the First and Second Observing Runs}",
    eprint = "1811.12907",
    archivePrefix = "arXiv",
    primaryClass = "astro-ph.HE",
    reportNumber = "LIGO-P1800307",
    doi = "10.1103/PhysRevX.9.031040",
    journal = "Phys. Rev. X",
    volume = "9",
    number = "3",
    pages = "031040",
    year = "2019"
}

@article{LIGOScientific:2020ibl,
    author = "Abbott, R. and others",
    collaboration = "LIGO Scientific, Virgo",
    title = "{GWTC-2: Compact Binary Coalescences Observed by LIGO and Virgo During the First Half of the Third Observing Run}",
    eprint = "2010.14527",
    archivePrefix = "arXiv",
    primaryClass = "gr-qc",
    reportNumber = "P2000061",
    doi = "10.1103/PhysRevX.11.021053",
    journal = "Phys. Rev. X",
    volume = "11",
    pages = "021053",
    year = "2021"
}

@article{LIGOScientific:2021usb,
    author = "Abbott, R. and others",
    collaboration = "LIGO Scientific, VIRGO",
    title = "{GWTC-2.1: Deep extended catalog of compact binary coalescences observed by LIGO and Virgo during the first half of the third observing run}",
    eprint = "2108.01045",
    archivePrefix = "arXiv",
    primaryClass = "gr-qc",
    reportNumber = "LIGO-P2100063",
    doi = "10.1103/PhysRevD.109.022001",
    journal = "Phys. Rev. D",
    volume = "109",
    number = "2",
    pages = "022001",
    year = "2024"
}

@article{KAGRA:2021vkt,
    author = "Abbott, R. and others",
    collaboration = "KAGRA, VIRGO, LIGO Scientific",
    title = "{GWTC-3: Compact Binary Coalescences Observed by LIGO and Virgo during the Second Part of the Third Observing Run}",
    eprint = "2111.03606",
    archivePrefix = "arXiv",
    primaryClass = "gr-qc",
    reportNumber = "LIGO-P2000318",
    doi = "10.1103/PhysRevX.13.041039",
    journal = "Phys. Rev. X",
    volume = "13",
    number = "4",
    pages = "041039",
    year = "2023"
}

@article{Bethe:1998bn,
    author = "Bethe, Hans A. and Brown, G. E.",
    title = "{Evolution of binary compact objects which merge}",
    eprint = "astro-ph/9802084",
    archivePrefix = "arXiv",
    reportNumber = "SUNY-NTG-98-4",
    doi = "10.1086/306265",
    journal = "Astrophys. J.",
    volume = "506",
    pages = "780--789",
    year = "1998"
}

@article{Kowalska:2010qg,
    author = "Kowalska, I. and Bulik, T. and Belczynski, K. and Dominik, M. and Gondek-Rosinska, D.",
    title = "{The eccentricity distribution of compact binaries}",
    eprint = "1010.0511",
    archivePrefix = "arXiv",
    primaryClass = "astro-ph.CO",
    doi = "10.1051/0004-6361/201015777",
    journal = "Astron. Astrophys.",
    volume = "527",
    pages = "A70",
    year = "2011"
}

@article{Stevenson:2017tfq,
    author = "Stevenson, Simon and Vigna-G{\'o}mez, Alejandro and Mandel, Ilya and Barrett, Jim W. and Neijssel, Coenraad J. and Perkins, David and de Mink, Selma E.",
    title = "{Formation of the first three gravitational-wave observations through isolated binary evolution}",
    eprint = "1704.01352",
    archivePrefix = "arXiv",
    primaryClass = "astro-ph.HE",
    doi = "10.1038/ncomms14906",
    journal = "Nature Commun.",
    volume = "8",
    pages = "14906",
    year = "2017"
}

@article{Bouffanais:2021wcr,
    author = "Bouffanais, Yann and Mapelli, Michela and Santoliquido, Filippo and Giacobbo, Nicola and Di Carlo, Ugo N. and Rastello, Sara and Artale, M. Celeste and Iorio, Giuliano",
    title = "{New insights on binary black hole formation channels after GWTC-2: young star clusters versus isolated binaries}",
    eprint = "2102.12495",
    archivePrefix = "arXiv",
    primaryClass = "astro-ph.HE",
    doi = "10.1093/mnras/stab2438",
    journal = "Mon. Not. Roy. Astron. Soc.",
    volume = "507",
    number = "4",
    pages = "5224--5235",
    year = "2021"
}

@article{KAGRA:2021duu,
    author = "Abbott, R. and others",
    collaboration = "KAGRA, VIRGO, LIGO Scientific",
    title = "{Population of Merging Compact Binaries Inferred Using Gravitational Waves through GWTC-3}",
    eprint = "2111.03634",
    archivePrefix = "arXiv",
    primaryClass = "astro-ph.HE",
    reportNumber = "LIGO-P2100239 ; Data release: https://zenodo.org/record/5655785, LIGO-P2100239",
    doi = "10.1103/PhysRevX.13.011048",
    journal = "Phys. Rev. X",
    volume = "13",
    number = "1",
    pages = "011048",
    year = "2023"
}

@article{Rodriguez:2016vmx,
    author = "Rodriguez, Carl L. and Zevin, Michael and Pankow, Chris and Kalogera, Vasilliki and Rasio, Frederic A.",
    title = "{Illuminating Black Hole Binary Formation Channels with Spins in Advanced LIGO}",
    eprint = "1609.05916",
    archivePrefix = "arXiv",
    primaryClass = "astro-ph.HE",
    doi = "10.3847/2041-8205/832/1/L2",
    journal = "Astrophys. J. Lett.",
    volume = "832",
    number = "1",
    pages = "L2",
    year = "2016"
}

@article{Belczynski:2014iua,
    author = "Belczynski, Krzysztof and Buonanno, Alessandra and Cantiello, Matteo and Fryer, Chris L. and Holz, Daniel E. and Mandel, Ilya and Miller, M. Coleman and Walczak, Marek",
    title = "{The Formation and Gravitational-Wave Detection of Massive Stellar Black-Hole Binaries}",
    eprint = "1403.0677",
    archivePrefix = "arXiv",
    primaryClass = "astro-ph.HE",
    reportNumber = "LA-UR-NUMBER-'LA-UR-14-21051'",
    doi = "10.1088/0004-637X/789/2/120",
    journal = "Astrophys. J.",
    volume = "789",
    number = "2",
    pages = "120",
    year = "2014"
}

@article{Mandel:2009nx,
    author = "Mandel, Ilya and O'Shaughnessy, Richard",
    editor = "Husa, Sascha and Krishnan, Badri",
    title = "{Compact Binary Coalescences in the Band of Ground-based Gravitational-Wave Detectors}",
    eprint = "0912.1074",
    archivePrefix = "arXiv",
    primaryClass = "astro-ph.HE",
    doi = "10.1088/0264-9381/27/11/114007",
    journal = "Class. Quant. Grav.",
    volume = "27",
    pages = "114007",
    year = "2010"
}

@article{Peters:1964zz,
    author = "Peters, P. C.",
    title = "{Gravitational Radiation and the Motion of Two Point Masses}",
    doi = "10.1103/PhysRev.136.B1224",
    journal = "Phys. Rev.",
    volume = "136",
    pages = "B1224--B1232",
    year = "1964"
}

@article{Divyajyoti:2023rht,
    author = "Divyajyoti and Kumar, Sumit and Tibrewal, Snehal and Romero-Shaw, Isobel M. and Mishra, Chandra Kant",
    title = "{Blind spots and biases: The dangers of ignoring eccentricity in gravitational-wave signals from binary black holes}",
    eprint = "2309.16638",
    archivePrefix = "arXiv",
    primaryClass = "gr-qc",
    doi = "10.1103/PhysRevD.109.043037",
    journal = "Phys. Rev. D",
    volume = "109",
    number = "4",
    pages = "043037",
    year = "2024"
}

@article{GilChoi:2022waq,
    author = "Gil Choi, Han and Yang, Tao and Lee, Hyung Mok",
    title = "{Importance of eccentricities in parameter estimation of compact binary inspirals with decihertz gravitational-wave detectors}",
    eprint = "2210.09541",
    archivePrefix = "arXiv",
    primaryClass = "gr-qc",
    doi = "10.1103/PhysRevD.110.024025",
    journal = "Phys. Rev. D",
    volume = "110",
    number = "2",
    pages = "024025",
    year = "2024"
}

@article{Cutler:2007mi,
    author = "Cutler, Curt and Vallisneri, Michele",
    title = "{LISA detections of massive black hole inspirals: Parameter extraction errors due to inaccurate template waveforms}",
    eprint = "0707.2982",
    archivePrefix = "arXiv",
    primaryClass = "gr-qc",
    doi = "10.1103/PhysRevD.76.104018",
    journal = "Phys. Rev. D",
    volume = "76",
    pages = "104018",
    year = "2007"
}

@article{Gayathri:2020coq,
    author = "Gayathri, V. and Healy, J. and Lange, J. and O'Brien, B. and Szczepanczyk, M. and Bartos, Imre and Campanelli, M. and Klimenko, S. and Lousto, C. O. and O'Shaughnessy, R.",
    title = "{Eccentricity estimate for black hole mergers with numerical relativity simulations}",
    eprint = "2009.05461",
    archivePrefix = "arXiv",
    primaryClass = "astro-ph.HE",
    doi = "10.1038/s41550-021-01568-w",
    journal = "Nature Astron.",
    volume = "6",
    number = "3",
    pages = "344--349",
    year = "2022"
}

@article{Yang:2023zxk,
    author = "Yang, Tao and Cai, Rong-Gen and Cao, Zhoujian and Lee, Hyung Mok",
    title = "{Eccentricity enables the earliest warning and localization of gravitational waves with ground-based detectors}",
    eprint = "2310.08160",
    archivePrefix = "arXiv",
    primaryClass = "gr-qc",
    doi = "10.1103/PhysRevD.109.104041",
    journal = "Phys. Rev. D",
    volume = "109",
    number = "10",
    pages = "104041",
    year = "2024"
}

@article{Zevin:2021rtf,
    author = "Zevin, Michael and Romero-Shaw, Isobel M. and Kremer, Kyle and Thrane, Eric and Lasky, Paul D.",
    title = "{Implications of Eccentric Observations on Binary Black Hole Formation Channels}",
    eprint = "2106.09042",
    archivePrefix = "arXiv",
    primaryClass = "astro-ph.HE",
    doi = "10.3847/2041-8213/ac32dc",
    journal = "Astrophys. J. Lett.",
    volume = "921",
    number = "2",
    pages = "L43",
    year = "2021"
}

@article{OLeary:2008myb,
    author = "O'Leary, Ryan M. and Kocsis, Bence and Loeb, Abraham",
    title = "{Gravitational waves from scattering of stellar-mass black holes in galactic nuclei}",
    eprint = "0807.2638",
    archivePrefix = "arXiv",
    primaryClass = "astro-ph",
    doi = "10.1111/j.1365-2966.2009.14653.x",
    journal = "Mon. Not. Roy. Astron. Soc.",
    volume = "395",
    number = "4",
    pages = "2127--2146",
    year = "2009"
}

@article{Samsing:2013kua,
    author = "Samsing, Johan and MacLeod, Morgan and Ramirez-Ruiz, Enrico",
    title = "{The Formation of Eccentric Compact Binary Inspirals and the Role of Gravitational Wave Emission in Binary-Single Stellar Encounters}",
    eprint = "1308.2964",
    archivePrefix = "arXiv",
    primaryClass = "astro-ph.HE",
    doi = "10.1088/0004-637X/784/1/71",
    journal = "Astrophys. J.",
    volume = "784",
    pages = "71",
    year = "2014"
}

@article{Samsing:2017rat,
    author = "Samsing, Johan and Ramirez-Ruiz, Enrico",
    title = "{On the Assembly Rate of Highly Eccentric Binary Black Hole Mergers}",
    eprint = "1703.09703",
    archivePrefix = "arXiv",
    primaryClass = "astro-ph.HE",
    doi = "10.3847/2041-8213/aa6f0b",
    journal = "Astrophys. J. Lett.",
    volume = "840",
    number = "2",
    pages = "L14",
    year = "2017"
}

@article{Gondan:2017wzd,
    author = "Gond{\'a}n, L{\'a}szl{\'o} and Kocsis, Bence and Raffai, P{\'e}ter and Frei, Zsolt",
    title = "{Eccentric Black Hole Gravitational-Wave Capture Sources in Galactic Nuclei: Distribution of Binary Parameters}",
    eprint = "1711.09989",
    archivePrefix = "arXiv",
    primaryClass = "astro-ph.HE",
    doi = "10.3847/1538-4357/aabfee",
    journal = "Astrophys. J.",
    volume = "860",
    number = "1",
    pages = "5",
    year = "2018"
}

@article{Kocsis:2011jy,
    author = "Kocsis, Bence and Levin, Janna",
    title = "{Repeated Bursts from Relativistic Scattering of Compact Objects in Galactic Nuclei}",
    eprint = "1109.4170",
    archivePrefix = "arXiv",
    primaryClass = "astro-ph.CO",
    doi = "10.1103/PhysRevD.85.123005",
    journal = "Phys. Rev. D",
    volume = "85",
    pages = "123005",
    year = "2012"
}

@article{Michaely:2019aet,
    author = "Michaely, Erez and Perets, Hagai B.",
    title = "{Gravitational-wave Sources from Mergers of Binary Black Holes Catalyzed by Flyby Interactions in the Field}",
    eprint = "1902.01864",
    archivePrefix = "arXiv",
    primaryClass = "astro-ph.SR",
    doi = "10.3847/2041-8213/ab5b9b",
    journal = "Astrophys. J. Lett.",
    volume = "887",
    number = "2",
    pages = "L36",
    year = "2019"
}

@article{LIGOScientific:2023lpe,
    author = "Abac, A. G. and others",
    collaboration = "LIGO Scientific, KAGRA, VIRGO",
    title = "{Search for Eccentric Black Hole Coalescences during the Third Observing Run of LIGO and Virgo}",
    eprint = "2308.03822",
    archivePrefix = "arXiv",
    primaryClass = "astro-ph.HE",
    reportNumber = "LIGO-P2300080",
    doi = "10.3847/1538-4357/ad65ce",
    journal = "Astrophys. J.",
    volume = "973",
    number = "2",
    pages = "132",
    year = "2024"
}

@article{LIGOScientific:2019dag,
    author = "Abbott, B. P. and others",
    collaboration = "LIGO Scientific, Virgo",
    title = "{Search for Eccentric Binary Black Hole Mergers with Advanced LIGO and Advanced Virgo during their First and Second Observing Runs}",
    eprint = "1907.09384",
    archivePrefix = "arXiv",
    primaryClass = "astro-ph.HE",
    reportNumber = "LIGO Document P1900110",
    doi = "10.3847/1538-4357/ab3c2d",
    journal = "Astrophys. J.",
    volume = "883",
    number = "2",
    pages = "149",
    year = "2019"
}

@article{Romero-Shaw:2020thy,
    author = "Romero-Shaw, Isobel M. and Lasky, Paul D. and Thrane, Eric and Bustillo, Juan Calderon",
    title = "{GW190521: orbital eccentricity and signatures of dynamical formation in a binary black hole merger signal}",
    eprint = "2009.04771",
    archivePrefix = "arXiv",
    primaryClass = "astro-ph.HE",
    doi = "10.3847/2041-8213/abbe26",
    journal = "Astrophys. J. Lett.",
    volume = "903",
    number = "1",
    pages = "L5",
    year = "2020"
}

@article{Romero-Shaw:2021ual,
    author = "Romero-Shaw, Isobel M. and Lasky, Paul D. and Thrane, Eric",
    title = "{Signs of Eccentricity in Two Gravitational-wave Signals May Indicate a Subpopulation of Dynamically Assembled Binary Black Holes}",
    eprint = "2108.01284",
    archivePrefix = "arXiv",
    primaryClass = "astro-ph.HE",
    doi = "10.3847/2041-8213/ac3138",
    journal = "Astrophys. J. Lett.",
    volume = "921",
    number = "2",
    pages = "L31",
    year = "2021"
}

@article{Romero-Shaw:2022xko,
    author = "Romero-Shaw, Isobel M. and Lasky, Paul D. and Thrane, Eric",
    title = "{Four Eccentric Mergers Increase the Evidence that LIGO{\textendash}Virgo{\textendash}KAGRA{\textquoteright}s Binary Black Holes Form Dynamically}",
    eprint = "2206.14695",
    archivePrefix = "arXiv",
    primaryClass = "astro-ph.HE",
    doi = "10.3847/1538-4357/ac9798",
    journal = "Astrophys. J.",
    volume = "940",
    number = "2",
    pages = "171",
    year = "2022"
}

@article{Gupte:2024jfe,
    author = "Gupte, Nihar and others",
    title = "{Evidence for eccentricity in the population of binary black holes observed by LIGO-Virgo-KAGRA}",
    eprint = "2404.14286",
    archivePrefix = "arXiv",
    primaryClass = "gr-qc",
    doi = "10.1103/vpyp-nvfp",
    journal = "Phys. Rev. D",
    volume = "112",
    number = "10",
    pages = "104045",
    year = "2025"
}

@article{Planas:2025jny,
    author = "Planas, Maria de Lluc and Ramos-Buades, Antoni and Garc{\'\i}a-Quir{\'o}s, Cecilio and Estell{\'e}s, H{\'e}ctor and Husa, Sascha and Haney, Maria",
    title = "{Reanalysis of binary black hole gravitational wave events for orbital eccentricity signatures}",
    eprint = "2504.15833",
    archivePrefix = "arXiv",
    primaryClass = "gr-qc",
    doi = "10.1103/cv75-y8dr",
    journal = "Phys. Rev. D",
    volume = "112",
    number = "12",
    pages = "123004",
    year = "2025"
}

@article{LIGOScientific:2020ufj,
    author = "Abbott, R. and others",
    collaboration = "LIGO Scientific, Virgo",
    title = "{Properties and Astrophysical Implications of the 150 M$_\odot$ Binary Black Hole Merger GW190521}",
    eprint = "2009.01190",
    archivePrefix = "arXiv",
    primaryClass = "astro-ph.HE",
    reportNumber = "LIGO-P2000021",
    doi = "10.3847/2041-8213/aba493",
    journal = "Astrophys. J. Lett.",
    volume = "900",
    number = "1",
    pages = "L13",
    year = "2020"
}

@article{Cao:2017ndf,
    author = "Cao, Zhoujian and Han, Wen-Biao",
    title = "{Waveform model for an eccentric binary black hole based on the effective-one-body-numerical-relativity formalism}",
    eprint = "1708.00166",
    archivePrefix = "arXiv",
    primaryClass = "gr-qc",
    doi = "10.1103/PhysRevD.96.044028",
    journal = "Phys. Rev. D",
    volume = "96",
    number = "4",
    pages = "044028",
    year = "2017"
}

@article{Liu:2019jpg,
    author = "Liu, Xiaolin and Cao, Zhoujian and Shao, Lijing",
    title = "{Validating the Effective-One-Body Numerical-Relativity Waveform Models for Spin-aligned Binary Black Holes along Eccentric Orbits}",
    eprint = "1910.00784",
    archivePrefix = "arXiv",
    primaryClass = "gr-qc",
    doi = "10.1103/PhysRevD.101.044049",
    journal = "Phys. Rev. D",
    volume = "101",
    number = "4",
    pages = "044049",
    year = "2020"
}

@article{Iglesias:2022xfc,
    author = "Iglesias, H. L. and others",
    title = "{Eccentricity Estimation for Five Binary Black Hole Mergers with Higher-order Gravitational-wave Modes}",
    eprint = "2208.01766",
    archivePrefix = "arXiv",
    primaryClass = "gr-qc",
    reportNumber = "LIGO-P2200208",
    doi = "10.3847/1538-4357/ad5ff6",
    journal = "Astrophys. J.",
    volume = "972",
    number = "1",
    pages = "65",
    year = "2024"
}

@article{Varma:2014jxa,
    author = {Varma, Vijay and Ajith, Parameswaran and Husa, Sascha and Bustillo, Juan Calderon and Hannam, Mark and P{\"u}rrer, Michael},
    title = "{Gravitational-wave observations of binary black holes: Effect of nonquadrupole modes}",
    eprint = "1409.2349",
    archivePrefix = "arXiv",
    primaryClass = "gr-qc",
    reportNumber = "LIGO-P1400095-V3",
    doi = "10.1103/PhysRevD.90.124004",
    journal = "Phys. Rev. D",
    volume = "90",
    number = "12",
    pages = "124004",
    year = "2014"
}

@article{Pekowsky:2012sr,
    author = "Pekowsky, Larne and Healy, James and Shoemaker, Deirdre and Laguna, Pablo",
    title = "{Impact of higher-order modes on the detection of binary black hole coalescences}",
    eprint = "1210.1891",
    archivePrefix = "arXiv",
    primaryClass = "gr-qc",
    reportNumber = "NSF-KITP-12-174",
    doi = "10.1103/PhysRevD.87.084008",
    journal = "Phys. Rev. D",
    volume = "87",
    number = "8",
    pages = "084008",
    year = "2013"
}

@article{Ramos-Buades:2023yhy,
    author = "Ramos-Buades, Antoni and Buonanno, Alessandra and Gair, Jonathan",
    title = "{Bayesian inference of binary black holes with inspiral-merger-ringdown waveforms using two eccentric parameters}",
    eprint = "2309.15528",
    archivePrefix = "arXiv",
    primaryClass = "gr-qc",
    doi = "10.1103/PhysRevD.108.124063",
    journal = "Phys. Rev. D",
    volume = "108",
    number = "12",
    pages = "124063",
    year = "2023"
}

@article{Nagar:2024dzj,
    author = "Nagar, Alessandro and Gamba, Rossella and Rettegno, Piero and Fantini, Veronica and Bernuzzi, Sebastiano",
    title = "{Effective-one-body waveform model for noncircularized, planar, coalescing black hole binaries: The importance of radiation reaction}",
    eprint = "2404.05288",
    archivePrefix = "arXiv",
    primaryClass = "gr-qc",
    doi = "10.1103/PhysRevD.110.084001",
    journal = "Phys. Rev. D",
    volume = "110",
    number = "8",
    pages = "084001",
    year = "2024"
}

@article{Chandramouli:2024vhw,
    author = "Chandramouli, Rohit S. and Prokup, Kaitlyn and Berti, Emanuele and Yunes, Nicol{\'a}s",
    title = "{Systematic biases due to waveform mismodeling in parametrized post-Einsteinian tests of general relativity: The impact of neglecting spin precession and higher modes}",
    eprint = "2410.06254",
    archivePrefix = "arXiv",
    primaryClass = "gr-qc",
    doi = "10.1103/PhysRevD.111.044026",
    journal = "Phys. Rev. D",
    volume = "111",
    number = "4",
    pages = "044026",
    year = "2025"
}

@article{Yi:2025pxe,
    author = "Yi, Sophia and Iacovelli, Francesco and Marsat, Sylvain and Wadekar, Digvijay and Berti, Emanuele",
    title = "{Systematic biases in parameter estimation on LISA binaries: The effect of excluding higher harmonics for non-spinning binaries}",
    eprint = "2502.12237",
    archivePrefix = "arXiv",
    primaryClass = "gr-qc",
    doi = "10.1103/3gs3-gmb4",
    journal = "Phys. Rev. D",
    volume = "112",
    number = "12",
    pages = "124063",
    year = "2025"
}

@article{Yunes:2009yz,
    author = "Yunes, Nicolas and Arun, K. G. and Berti, Emanuele and Will, Clifford M.",
    title = "{Post-Circular Expansion of Eccentric Binary Inspirals: Fourier-Domain Waveforms in the Stationary Phase Approximation}",
    eprint = "0906.0313",
    archivePrefix = "arXiv",
    primaryClass = "gr-qc",
    doi = "10.1103/PhysRevD.80.084001",
    journal = "Phys. Rev. D",
    volume = "80",
    number = "8",
    pages = "084001",
    year = "2009",
    note = "[Erratum: Phys.Rev.D 89, 109901 (2014)]"
}

@article{Cornish:2010cd,
    author = "Cornish, Neil J. and Shapiro Key, Joey",
    title = "{Computing waveforms for spinning compact binaries in quasi-eccentric orbits}",
    eprint = "1004.5322",
    archivePrefix = "arXiv",
    primaryClass = "gr-qc",
    doi = "10.1103/PhysRevD.82.044028",
    journal = "Phys. Rev. D",
    volume = "82",
    pages = "044028",
    year = "2010",
    note = "[Erratum: Phys.Rev.D 84, 029901 (2011)]"
}

@article{ShapiroKey:2010cnz,
    author = "Shapiro Key, Joey and Cornish, Neil J.",
    title = "{Characterizing Spinning Black Hole Binaries in Eccentric Orbits with LISA}",
    eprint = "1006.3759",
    archivePrefix = "arXiv",
    primaryClass = "gr-qc",
    doi = "10.1103/PhysRevD.83.083001",
    journal = "Phys. Rev. D",
    volume = "83",
    pages = "083001",
    year = "2011"
}

@article{Huerta:2014eca,
    author = "Huerta, E. A. and Kumar, Prayush and McWilliams, Sean T. and O'Shaughnessy, Richard and Yunes, Nicol{\'a}s",
    title = "{Accurate and efficient waveforms for compact binaries on eccentric orbits}",
    eprint = "1408.3406",
    archivePrefix = "arXiv",
    primaryClass = "gr-qc",
    doi = "10.1103/PhysRevD.90.084016",
    journal = "Phys. Rev. D",
    volume = "90",
    number = "8",
    pages = "084016",
    year = "2014"
}

@article{Loutrel:2017fgu,
    author = "Loutrel, Nicholas and Yunes, Nicol{\'a}s",
    title = "{Eccentric Gravitational Wave Bursts in the Post-Newtonian Formalism}",
    eprint = "1702.01818",
    archivePrefix = "arXiv",
    primaryClass = "gr-qc",
    doi = "10.1088/1361-6382/aa7449",
    journal = "Class. Quant. Grav.",
    volume = "34",
    number = "13",
    pages = "135011",
    year = "2017"
}

@article{Klein:2018ybm,
    author = "Klein, Antoine and Boetzel, Yannick and Gopakumar, Achamveedu and Jetzer, Philippe and de Vittori, Lorenzo",
    title = "{Fourier domain gravitational waveforms for precessing eccentric binaries}",
    eprint = "1801.08542",
    archivePrefix = "arXiv",
    primaryClass = "gr-qc",
    doi = "10.1103/PhysRevD.98.104043",
    journal = "Phys. Rev. D",
    volume = "98",
    number = "10",
    pages = "104043",
    year = "2018"
}

@misc{klein2021efpeefficientfullyprecessing,
      title={EFPE: Efficient fully precessing eccentric gravitational waveforms for binaries with long inspirals}, 
      author={Antoine Klein},
      year={2021},
      eprint={2106.10291},
      archivePrefix={arXiv},
      primaryClass={gr-qc},
      url={https://arxiv.org/abs/2106.10291}, 
}

@article{Tanay:2016zog,
    author = "Tanay, Sashwat and Haney, Maria and Gopakumar, Achamveedu",
    title = "{Frequency and time domain inspiral templates for comparable mass compact binaries in eccentric orbits}",
    eprint = "1602.03081",
    archivePrefix = "arXiv",
    primaryClass = "gr-qc",
    doi = "10.1103/PhysRevD.93.064031",
    journal = "Phys. Rev. D",
    volume = "93",
    number = "6",
    pages = "064031",
    year = "2016"
}

@article{Moore:2016qxz,
    author = "Moore, Blake and Favata, Marc and Arun, K. G. and Mishra, Chandra Kant",
    title = "{Gravitational-wave phasing for low-eccentricity inspiralling compact binaries to 3PN order}",
    eprint = "1605.00304",
    archivePrefix = "arXiv",
    primaryClass = "gr-qc",
    reportNumber = "LIGO-DCC-P1500268",
    doi = "10.1103/PhysRevD.93.124061",
    journal = "Phys. Rev. D",
    volume = "93",
    number = "12",
    pages = "124061",
    year = "2016"
}

@article{Moore:2018kvz,
    author = "Moore, Blake and Robson, Travis and Loutrel, Nicholas and Yunes, Nicolas",
    title = "{Towards a Fourier domain waveform for non-spinning binaries with arbitrary eccentricity}",
    eprint = "1807.07163",
    archivePrefix = "arXiv",
    primaryClass = "gr-qc",
    doi = "10.1088/1361-6382/aaea00",
    journal = "Class. Quant. Grav.",
    volume = "35",
    number = "23",
    pages = "235006",
    year = "2018"
}

@article{Moore:2019xkm,
    author = "Moore, Blake and Yunes, Nicol{\'a}s",
    title = "{A 3PN Fourier Domain Waveform for Non-Spinning Binaries with Moderate Eccentricity}",
    eprint = "1903.05203",
    archivePrefix = "arXiv",
    primaryClass = "gr-qc",
    doi = "10.1088/1361-6382/ab3778",
    journal = "Class. Quant. Grav.",
    volume = "36",
    number = "18",
    pages = "185003",
    year = "2019"
}

@article{Islam:2021mha,
    author = "Islam, Tousif and Varma, Vijay and Lodman, Jackie and Field, Scott E. and Khanna, Gaurav and Scheel, Mark A. and Pfeiffer, Harald P. and Gerosa, Davide and Kidder, Lawrence E.",
    title = "{Eccentric binary black hole surrogate models for the gravitational waveform and remnant properties: comparable mass, nonspinning case}",
    eprint = "2101.11798",
    archivePrefix = "arXiv",
    primaryClass = "gr-qc",
    doi = "10.1103/PhysRevD.103.064022",
    journal = "Phys. Rev. D",
    volume = "103",
    number = "6",
    pages = "064022",
    year = "2021"
}

@article{Nagar:2021gss,
    author = "Nagar, Alessandro and Bonino, Alice and Rettegno, Piero",
    title = "{Effective one-body multipolar waveform model for spin-aligned, quasicircular, eccentric, hyperbolic black hole binaries}",
    eprint = "2101.08624",
    archivePrefix = "arXiv",
    primaryClass = "gr-qc",
    doi = "10.1103/PhysRevD.103.104021",
    journal = "Phys. Rev. D",
    volume = "103",
    number = "10",
    pages = "104021",
    year = "2021"
}

@article{Liu:2021pkr,
    author = "Liu, Xiaolin and Cao, Zhoujian and Zhu, Zong-Hong",
    title = "{A higher-multipole gravitational waveform model for an eccentric binary black holes based on the effective-one-body-numerical-relativity formalism}",
    eprint = "2102.08614",
    archivePrefix = "arXiv",
    primaryClass = "gr-qc",
    doi = "10.1088/1361-6382/ac4119",
    journal = "Class. Quant. Grav.",
    volume = "39",
    number = "3",
    pages = "035009",
    year = "2022"
}

@article{Ramos-Buades:2021adz,
    author = "Ramos-Buades, Antoni and Buonanno, Alessandra and Khalil, Mohammed and Ossokine, Serguei",
    title = "{Effective-one-body multipolar waveforms for eccentric binary black holes with nonprecessing spins}",
    eprint = "2112.06952",
    archivePrefix = "arXiv",
    primaryClass = "gr-qc",
    doi = "10.1103/PhysRevD.105.044035",
    journal = "Phys. Rev. D",
    volume = "105",
    number = "4",
    pages = "044035",
    year = "2022"
}

@article{Liu:2023ldr,
    author = "Liu, Xiaolin and Cao, Zhoujian and Zhu, Zong-Hong",
    title = "{Effective-one-body numerical-relativity waveform model for eccentric spin-precessing binary black hole coalescence}",
    eprint = "2310.04552",
    archivePrefix = "arXiv",
    primaryClass = "gr-qc",
    doi = "10.1088/1361-6382/ad72ca",
    journal = "Class. Quant. Grav.",
    volume = "41",
    number = "19",
    pages = "195019",
    year = "2024"
}

@article{Gamboa:2024hli,
    author = "Gamboa, Aldo and others",
    title = "{Accurate waveforms for eccentric, aligned-spin binary black holes: The multipolar effective-one-body model seobnrv5ehm}",
    eprint = "2412.12823",
    archivePrefix = "arXiv",
    primaryClass = "gr-qc",
    doi = "10.1103/jxrc-z298",
    journal = "Phys. Rev. D",
    volume = "112",
    number = "4",
    pages = "044038",
    year = "2025"
}

@article{Pompili:2023tna,
    author = "Pompili, Lorenzo and others",
    title = "{Laying the foundation of the effective-one-body waveform models SEOBNRv5: Improved accuracy and efficiency for spinning nonprecessing binary black holes}",
    eprint = "2303.18039",
    archivePrefix = "arXiv",
    primaryClass = "gr-qc",
    doi = "10.1103/PhysRevD.108.124035",
    journal = "Phys. Rev. D",
    volume = "108",
    number = "12",
    pages = "124035",
    year = "2023"
}

@article{Planas:2025feq,
    author = "Planas, Maria de Lluc and Ramos-Buades, Antoni and Garc{\'\i}a-Quir{\'o}s, Cecilio and Estell{\'e}s, H{\'e}ctor and Husa, Sascha and Haney, Maria",
    title = "{Time-domain phenomenological multipolar waveforms for aligned-spin binary black holes in elliptical orbits}",
    eprint = "2503.13062",
    archivePrefix = "arXiv",
    primaryClass = "gr-qc",
    doi = "10.1103/wz3v-b151",
    journal = "Phys. Rev. D",
    volume = "113",
    number = "2",
    pages = "024006",
    year = "2026"
}

@article{bilby_paper,
    author = "Ashton, Gregory and others",
    title = "{BILBY: A user-friendly Bayesian inference library for gravitational-wave astronomy}",
    eprint = "1811.02042",
    archivePrefix = "arXiv",
    primaryClass = "astro-ph.IM",
    doi = "10.3847/1538-4365/ab06fc",
    journal = "Astrophys. J. Suppl.",
    volume = "241",
    number = "2",
    pages = "27",
    year = "2019"
}

@article{Thrane:2018qnx,
    author = "Thrane, Eric and Talbot, Colm",
    title = "{An introduction to Bayesian inference in gravitational-wave astronomy: parameter estimation, model selection, and hierarchical models}",
    eprint = "1809.02293",
    archivePrefix = "arXiv",
    primaryClass = "astro-ph.IM",
    doi = "10.1017/pasa.2019.2",
    journal = "Publ. Astron. Soc. Austral.",
    volume = "36",
    pages = "e010",
    year = "2019",
    note = "[Erratum: Publ.Astron.Soc.Austral. 37, e036 (2020)]"
}

@article{Romero-Shaw:2020owr,
    author = "Romero-Shaw, I. M. and others",
    title = "{Bayesian inference for compact binary coalescences with bilby: validation and application to the first LIGO{\textendash}Virgo gravitational-wave transient catalogue}",
    eprint = "2006.00714",
    archivePrefix = "arXiv",
    primaryClass = "astro-ph.IM",
    doi = "10.1093/mnras/staa2850",
    journal = "Mon. Not. Roy. Astron. Soc.",
    volume = "499",
    number = "3",
    pages = "3295--3319",
    year = "2020"
}

@article{Speagle:2019ivv,
    author = "Speagle, Joshua S.",
    title = "{dynesty: a dynamic nested sampling package for estimating Bayesian posteriors and evidences}",
    eprint = "1904.02180",
    archivePrefix = "arXiv",
    primaryClass = "astro-ph.IM",
    doi = "10.1093/mnras/staa278",
    journal = "Mon. Not. Roy. Astron. Soc.",
    volume = "493",
    number = "3",
    pages = "3132--3158",
    year = "2020"
}

@article{Huerta:2016rwp,
    author = "Huerta, E. A. and others",
    title = "{Complete waveform model for compact binaries on eccentric orbits}",
    eprint = "1609.05933",
    archivePrefix = "arXiv",
    primaryClass = "gr-qc",
    doi = "10.1103/PhysRevD.95.024038",
    journal = "Phys. Rev. D",
    volume = "95",
    number = "2",
    pages = "024038",
    year = "2017"
}

@article{Huerta:2017kez,
    author = "Huerta, E. A. and others",
    title = "{Eccentric, nonspinning, inspiral, Gaussian-process merger approximant for the detection and characterization of eccentric binary black hole mergers}",
    eprint = "1711.06276",
    archivePrefix = "arXiv",
    primaryClass = "gr-qc",
    doi = "10.1103/PhysRevD.97.024031",
    journal = "Phys. Rev. D",
    volume = "97",
    number = "2",
    pages = "024031",
    year = "2018"
}

@article{Hinder:2017sxy,
    author = "Hinder, Ian and Kidder, Lawrence E. and Pfeiffer, Harald P.",
    title = "{Eccentric binary black hole inspiral-merger-ringdown gravitational waveform model from numerical relativity and post-Newtonian theory}",
    eprint = "1709.02007",
    archivePrefix = "arXiv",
    primaryClass = "gr-qc",
    doi = "10.1103/PhysRevD.98.044015",
    journal = "Phys. Rev. D",
    volume = "98",
    number = "4",
    pages = "044015",
    year = "2018"
}

@article{Chattaraj:2022tay,
    author = "Chattaraj, Abhishek and RoyChowdhury, Tamal and Divyajyoti and Mishra, Chandra Kant and Gupta, Anshu",
    title = "{High accuracy post-Newtonian and numerical relativity comparisons involving higher modes for eccentric binary black holes and a dominant mode eccentric inspiral-merger-ringdown model}",
    eprint = "2204.02377",
    archivePrefix = "arXiv",
    primaryClass = "gr-qc",
    reportNumber = "LIGO-P2200106",
    doi = "10.1103/PhysRevD.106.124008",
    journal = "Phys. Rev. D",
    volume = "106",
    number = "12",
    pages = "124008",
    year = "2022"
}

@article{Manna:2024ycx,
    author = "Manna, Pratul and RoyChowdhury, Tamal and Mishra, Chandra Kant",
    title = "{Improved inspiral-merger-ringdown model for BBHs on elliptical orbits}",
    eprint = "2409.10672",
    archivePrefix = "arXiv",
    primaryClass = "gr-qc",
    doi = "10.1103/849s-3zy8",
    journal = "Phys. Rev. D",
    volume = "111",
    number = "12",
    pages = "124026",
    year = "2025"
}

@article{Paul:2024ujx,
    author = "Paul, Kaushik and Maurya, Akash and Henry, Quentin and Sharma, Kartikey and Satheesh, Pranav and Divyajyoti and Kumar, Prayush and Mishra, Chandra Kant",
    title = "{Eccentric, spinning, inspiral-merger-ringdown waveform model with higher modes for the detection and characterization of binary black holes}",
    eprint = "2409.13866",
    archivePrefix = "arXiv",
    primaryClass = "gr-qc",
    doi = "10.1103/PhysRevD.111.084074",
    journal = "Phys. Rev. D",
    volume = "111",
    number = "8",
    pages = "084074",
    year = "2025"
}

@article{Ramos-Buades:2019uvh,
    author = "Ramos-Buades, Antoni and Husa, Sascha and Pratten, Geraint and Estell{\'e}s, H{\'e}ctor and Garc{\'\i}a-Quir{\'o}s, Cecilio and Mateu-Lucena, Maite and Colleoni, Marta and Jaume, Rafel",
    title = "{First survey of spinning eccentric black hole mergers: Numerical relativity simulations, hybrid waveforms, and parameter estimation}",
    eprint = "1909.11011",
    archivePrefix = "arXiv",
    primaryClass = "gr-qc",
    doi = "10.1103/PhysRevD.101.083015",
    journal = "Phys. Rev. D",
    volume = "101",
    number = "8",
    pages = "083015",
    year = "2020"
}

@article{Gadre:2024ndy,
    author = "Gadre, Bhooshan and Soni, Kanchan and Tiwari, Shubhanshu and Ramos-Buades, Antoni and Haney, Maria and Mitra, Sanjit",
    title = "{Detectability of eccentric binary black holes with matched filtering and unmodeled pipelines during the third observing run of LIGO-Virgo-KAGRA}",
    eprint = "2405.04186",
    archivePrefix = "arXiv",
    primaryClass = "gr-qc",
    reportNumber = "LIGO-P2400136, VIR-0327A-24",
    doi = "10.1103/PhysRevD.110.044013",
    journal = "Phys. Rev. D",
    volume = "110",
    number = "4",
    pages = "044013",
    year = "2024"
}

@article{Lin:1991zzm,
    author = "Lin, J.",
    title = "{Divergence measures based on the Shannon entropy}",
    doi = "10.1109/18.61115",
    journal = "IEEE Trans. Info. Theor.",
    volume = "37",
    number = "1",
    pages = "145--151",
    year = "1991"
}

@misc{LIGOScientific:2025slb,
    author = "Abac, A. G. and others",
    collaboration = "LIGO Scientific, VIRGO, KAGRA",
    title = "{GWTC-4.0: Updating the Gravitational-Wave Transient Catalog with Observations from the First Part of the Fourth LIGO-Virgo-KAGRA Observing Run}",
    eprint = "2508.18082",
    archivePrefix = "arXiv",
    primaryClass = "gr-qc",
    reportNumber = "LIGO-P2400386",
    month = "8",
    year = "2025"
}

@article{LIGOScientific:2016lio,
    author = "Abbott, B. P. and others",
    collaboration = "LIGO Scientific, Virgo",
    title = "{Tests of general relativity with GW150914}",
    eprint = "1602.03841",
    archivePrefix = "arXiv",
    primaryClass = "gr-qc",
    reportNumber = "LIGO-P1500213",
    doi = "10.1103/PhysRevLett.116.221101",
    journal = "Phys. Rev. Lett.",
    volume = "116",
    number = "22",
    pages = "221101",
    year = "2016",
    note = "[Erratum: Phys.Rev.Lett. 121, 129902 (2018)]"
}

@article{LIGOScientific:2025rid,
    author = "Abac, A. G. and others",
    collaboration = "LIGO Scientific, Virgo, KAGRA",
    title = "{GW250114: Testing Hawking{\textquoteright}s Area Law and the Kerr Nature of Black Holes}",
    eprint = "2509.08054",
    archivePrefix = "arXiv",
    primaryClass = "gr-qc",
    reportNumber = "LIGO-P2500421",
    doi = "10.1103/kw5g-d732",
    journal = "Phys. Rev. Lett.",
    volume = "135",
    number = "11",
    pages = "111403",
    year = "2025"
}

@article{Will:2014kxa,
    author = "Will, Clifford M.",
    title = "{The Confrontation between General Relativity and Experiment}",
    eprint = "1403.7377",
    archivePrefix = "arXiv",
    primaryClass = "gr-qc",
    doi = "10.12942/lrr-2014-4",
    journal = "Living Rev. Rel.",
    volume = "17",
    pages = "4",
    year = "2014"
}

@article{LIGOScientific:2018dkp,
    author = "Abbott, B. P. and others",
    collaboration = "LIGO Scientific, Virgo",
    title = "{Tests of General Relativity with GW170817}",
    eprint = "1811.00364",
    archivePrefix = "arXiv",
    primaryClass = "gr-qc",
    reportNumber = "LIGO-P1800059",
    doi = "10.1103/PhysRevLett.123.011102",
    journal = "Phys. Rev. Lett.",
    volume = "123",
    number = "1",
    pages = "011102",
    year = "2019"
}

@article{Krishnendu:2021fga,
    author = "Krishnendu, N. V. and Ohme, Frank",
    title = "{Testing General Relativity with Gravitational Waves: An Overview}",
    eprint = "2201.05418",
    archivePrefix = "arXiv",
    primaryClass = "gr-qc",
    doi = "10.3390/universe7120497",
    journal = "Universe",
    volume = "7",
    number = "12",
    pages = "497",
    year = "2021"
}

@misc{theligoscientificcollaboration2025blackholespectroscopytests,
      title={Black Hole Spectroscopy and Tests of General Relativity with GW250114}, 
      author={The LIGO Scientific Collaboration and the Virgo Collaboration and the KAGRA Collaboration},
      year={2025},
      eprint={2509.08099},
      archivePrefix={arXiv},
      primaryClass={gr-qc},
      url={https://arxiv.org/abs/2509.08099}, 
}

@article{LIGOScientific:2019fpa,
    author = "Abbott, B. P. and others",
    collaboration = "LIGO Scientific, Virgo",
    title = "{Tests of General Relativity with the Binary Black Hole Signals from the LIGO-Virgo Catalog GWTC-1}",
    eprint = "1903.04467",
    archivePrefix = "arXiv",
    primaryClass = "gr-qc",
    reportNumber = "LIGO-P1800316",
    doi = "10.1103/PhysRevD.100.104036",
    journal = "Phys. Rev. D",
    volume = "100",
    number = "10",
    pages = "104036",
    year = "2019"
}

@article{LIGOScientific:2020tif,
    author = "Abbott, R. and others",
    collaboration = "LIGO Scientific, Virgo",
    title = "{Tests of general relativity with binary black holes from the second LIGO-Virgo gravitational-wave transient catalog}",
    eprint = "2010.14529",
    archivePrefix = "arXiv",
    primaryClass = "gr-qc",
    reportNumber = "LIGO-P2000091",
    doi = "10.1103/PhysRevD.103.122002",
    journal = "Phys. Rev. D",
    volume = "103",
    number = "12",
    pages = "122002",
    year = "2021"
}

@misc{LIGOScientific:2021sio,
    author = "Abbott, R. and others",
    collaboration = "LIGO Scientific, VIRGO, KAGRA",
    title = "{Tests of General Relativity with GWTC-3}",
    eprint = "2112.06861",
    archivePrefix = "arXiv",
    primaryClass = "gr-qc",
    reportNumber = "LIGO-P2100275",
    month = "12",
    year = "2021"
}

@article{Schutz:1986gp,
    author = "Schutz, Bernard F.",
    title = "{Determining the Hubble Constant from Gravitational Wave Observations}",
    doi = "10.1038/323310a0",
    journal = "Nature",
    volume = "323",
    pages = "310--311",
    year = "1986"
}

@article{LIGOScientific:2017adf,
    author = "Abbott, B. P. and others",
    collaboration = "LIGO Scientific, Virgo, 1M2H, Dark Energy Camera GW-E, DES, DLT40, Las Cumbres Observatory, VINROUGE, MASTER",
    title = "{A gravitational-wave standard siren measurement of the Hubble constant}",
    eprint = "1710.05835",
    archivePrefix = "arXiv",
    primaryClass = "astro-ph.CO",
    reportNumber = "LIGO-P1700296, FERMILAB-PUB-17-472-A-AE",
    doi = "10.1038/nature24471",
    journal = "Nature",
    volume = "551",
    number = "7678",
    pages = "85--88",
    year = "2017"
}

@article{LIGOScientific:2018jsj,
    author = "Abbott, B. P. and others",
    collaboration = "LIGO Scientific, Virgo",
    title = "{Binary Black Hole Population Properties Inferred from the First and Second Observing Runs of Advanced LIGO and Advanced Virgo}",
    eprint = "1811.12940",
    archivePrefix = "arXiv",
    primaryClass = "astro-ph.HE",
    reportNumber = "LIGO-P1800324",
    doi = "10.3847/2041-8213/ab3800",
    journal = "Astrophys. J. Lett.",
    volume = "882",
    number = "2",
    pages = "L24",
    year = "2019"
}

@article{LIGOScientific:2020kqk,
    author = "Abbott, R. and others",
    collaboration = "LIGO Scientific, Virgo",
    title = "{Population Properties of Compact Objects from the Second LIGO-Virgo Gravitational-Wave Transient Catalog}",
    eprint = "2010.14533",
    archivePrefix = "arXiv",
    primaryClass = "astro-ph.HE",
    reportNumber = "LIGO-P2000077",
    doi = "10.3847/2041-8213/abe949",
    journal = "Astrophys. J. Lett.",
    volume = "913",
    number = "1",
    pages = "L7",
    year = "2021"
}

@misc{LIGOScientific:2025pvj,
    author = "Abac, A. G. and others",
    collaboration = "LIGO Scientific, VIRGO, KAGRA",
    title = "{GWTC-4.0: Population Properties of Merging Compact Binaries}",
    eprint = "2508.18083",
    archivePrefix = "arXiv",
    primaryClass = "astro-ph.HE",
    reportNumber = "LIGO-P2400004",
    month = "8",
    year = "2025"
}

@ARTICLE{1988ApJ...329..764L,
       author = {{Livio}, Mario and {Soker}, Noam},
        title = "{The Common Envelope Phase in the Evolution of Binary Stars}",
      journal = {\apj},
         year = 1988,
        month = jun,
       volume = {329},
        pages = {764},
          doi = {10.1086/166419},
       adsurl = {https://ui.adsabs.harvard.edu/abs/1988ApJ...329..764L}
}

@article{Ivanova:2012vx,
    author = "Ivanova, N. and others",
    title = "{Common Envelope Evolution: Where we stand and how we can move forward}",
    eprint = "1209.4302",
    archivePrefix = "arXiv",
    primaryClass = "astro-ph.HE",
    doi = "10.1007/s00159-013-0059-2",
    journal = "Astron. Astrophys. Rev.",
    volume = "21",
    pages = "59",
    year = "2013"
}

@article{Belczynski:2001uc,
    author = "Belczynski, Krzysztof and Kalogera, Vassiliki and Bulik, Tomasz",
    title = "{A Comprehensive study of binary compact objects as gravitational wave sources: Evolutionary channels, rates, and physical properties}",
    eprint = "astro-ph/0111452",
    archivePrefix = "arXiv",
    doi = "10.1086/340304",
    journal = "Astrophys. J.",
    volume = "572",
    pages = "407--431",
    year = "2002"
}

@article{Belczynski:2016obo,
    author = "Belczynski, Krzysztof and Holz, Daniel E. and Bulik, Tomasz and O'Shaughnessy, Richard",
    title = "{The first gravitational-wave source from the isolated evolution of two 40-100 Msun stars}",
    eprint = "1602.04531",
    archivePrefix = "arXiv",
    primaryClass = "astro-ph.HE",
    doi = "10.1038/nature18322",
    journal = "Nature",
    volume = "534",
    pages = "512",
    year = "2016"
}

@article{Marchant:2016wow,
    author = "Marchant, Pablo and Langer, Norbert and Podsiadlowski, Philipp and Tauris, Thomas M. and Moriya, Takashi J.",
    title = "{A new route towards merging massive black holes}",
    eprint = "1601.03718",
    archivePrefix = "arXiv",
    primaryClass = "astro-ph.SR",
    doi = "10.1051/0004-6361/201628133",
    journal = "Astron. Astrophys.",
    volume = "588",
    pages = "A50",
    year = "2016"
}

@article{Sigurdsson:1993zrm,
    author = "Sigurdsson, Steinn and Hernquist, Lars",
    title = "{Primordial black holes in globular clusters}",
    doi = "10.1038/364423a0",
    journal = "Nature",
    volume = "364",
    pages = "423--425",
    year = "1993"
}

@article{PortegiesZwart:1999nm,
    author = "Portegies Zwart, Simon F. and McMillan, Stephen",
    title = "{Black hole mergers in the universe}",
    eprint = "astro-ph/9910061",
    archivePrefix = "arXiv",
    doi = "10.1086/312422",
    journal = "Astrophys. J. Lett.",
    volume = "528",
    pages = "L17",
    year = "2000"
}

@article{OLeary:2005vqo,
    author = "O'Leary, Ryan M. and Rasio, Frederic A. and Fregeau, John M. and Ivanova, Natalia and O'Shaughnessy, Richard W.",
    title = "{Binary mergers and growth of black holes in dense star clusters}",
    eprint = "astro-ph/0508224",
    archivePrefix = "arXiv",
    doi = "10.1086/498446",
    journal = "Astrophys. J.",
    volume = "637",
    pages = "937--951",
    year = "2006"
}

@article{Zevin:2020gbd,
    author = "Zevin, Michael and Bavera, Simone S. and Berry, Christopher P. L. and Kalogera, Vicky and Fragos, Tassos and Marchant, Pablo and Rodriguez, Carl L. and Antonini, Fabio and Holz, Daniel E. and Pankow, Chris",
    title = "{One Channel to Rule Them All? Constraining the Origins of Binary Black Holes Using Multiple Formation Pathways}",
    eprint = "2011.10057",
    archivePrefix = "arXiv",
    primaryClass = "astro-ph.HE",
    doi = "10.3847/1538-4357/abe40e",
    journal = "Astrophys. J.",
    volume = "910",
    number = "2",
    pages = "152",
    year = "2021"
}

@article{Yang:2019okq,
    author = "Yang, Y. and Bartos, I. and Haiman, Z. and Kocsis, B. and Marka, Z. and Stone, N. C. and Marka, S.",
    title = "{AGN Disks Harden the Mass Distribution of Stellar-mass Binary Black Hole Mergers}",
    eprint = "1903.01405",
    archivePrefix = "arXiv",
    primaryClass = "astro-ph.HE",
    doi = "10.3847/1538-4357/ab16e3",
    journal = "Astrophys. J.",
    volume = "876",
    number = "2",
    pages = "122",
    year = "2019"
}

@article{Mckernan:2017ssq,
    author = "Mckernan, B. and others",
    title = "{Constraining Stellar-mass Black Hole Mergers in AGN Disks Detectable with LIGO}",
    eprint = "1702.07818",
    archivePrefix = "arXiv",
    primaryClass = "astro-ph.HE",
    doi = "10.3847/1538-4357/aadae5",
    journal = "Astrophys. J.",
    volume = "866",
    number = "1",
    pages = "66",
    year = "2018"
}

@article{Mapelli:2020vfa,
    author = "Mapelli, Michela",
    title = "{Binary Black Hole Mergers: Formation and Populations}",
    eprint = "2105.12455",
    archivePrefix = "arXiv",
    primaryClass = "astro-ph.HE",
    doi = "10.3389/fspas.2020.00038",
    journal = "Front. Astron. Space Sci.",
    volume = "7",
    pages = "38",
    year = "2020"
}

@article{Kozai:1962zz,
    author = "Kozai, Yoshihide",
    title = "{Secular perturbations of asteroids with high inclination and eccentricity}",
    doi = "10.1086/108790",
    journal = "Astron. J.",
    volume = "67",
    pages = "591--598",
    year = "1962"
}

@article{Lidov:1962wjn,
    author = "Lidov, M. L.",
    title = "{The evolution of orbits of artificial satellites of planets under the action of gravitational perturbations of external bodies}",
    doi = "10.1016/0032-0633(62)90129-0",
    journal = "Planet. Space Sci.",
    volume = "9",
    number = "10",
    pages = "719--759",
    year = "1962"
}

@article{VanLandingham:2016ccd,
    author = "VanLandingham, John H. and Miller, M. Coleman and Hamilton, Douglas P. and Richardson, Derek C.",
    title = "{The Role of the Kozai{\textendash}lidov Mechanism in Black Hole Binary Mergers in Galactic Centers}",
    eprint = "1604.04948",
    archivePrefix = "arXiv",
    primaryClass = "astro-ph.HE",
    doi = "10.3847/0004-637X/828/2/77",
    journal = "Astrophys. J.",
    volume = "828",
    number = "2",
    pages = "77",
    year = "2016"
}

@article{Antonini:2012ad,
    author = "Antonini, Fabio and Perets, Hagai B.",
    title = "{Secular evolution of compact binaries near massive black holes: Gravitational wave sources and other exotica}",
    eprint = "1203.2938",
    archivePrefix = "arXiv",
    primaryClass = "astro-ph.GA",
    doi = "10.1088/0004-637X/757/1/27",
    journal = "Astrophys. J.",
    volume = "757",
    pages = "27",
    year = "2012"
}

@article{Antonini:2017ash,
    author = "Antonini, Fabio and Toonen, Silvia and Hamers, Adrian S.",
    title = "{Binary black hole mergers from field triples: properties, rates and the impact of stellar evolution}",
    eprint = "1703.06614",
    archivePrefix = "arXiv",
    primaryClass = "astro-ph.GA",
    doi = "10.3847/1538-4357/aa6f5e",
    journal = "Astrophys. J.",
    volume = "841",
    number = "2",
    pages = "77",
    year = "2017"
}

@article{Antognini:2013lpa,
    author = "Antognini, Joe M. and Shappee, Benjamin J. and Thompson, Todd A. and Amaro-Seoane, Pau",
    title = "{Rapid Eccentricity Oscillations and the Mergers of Compact Objects in Hierarchical Triples}",
    eprint = "1308.5682",
    archivePrefix = "arXiv",
    primaryClass = "astro-ph.HE",
    doi = "10.1093/mnras/stu039",
    journal = "Mon. Not. Roy. Astron. Soc.",
    volume = "439",
    number = "1",
    pages = "1079--1091",
    year = "2014"
}

@article{Samsing:2017xmd,
    author = "Samsing, Johan",
    title = "{Eccentric Black Hole Mergers Forming in Globular Clusters}",
    eprint = "1711.07452",
    archivePrefix = "arXiv",
    primaryClass = "astro-ph.HE",
    doi = "10.1103/PhysRevD.97.103014",
    journal = "Phys. Rev. D",
    volume = "97",
    number = "10",
    pages = "103014",
    year = "2018"
}

@article{Gondan:2018khr,
    author = "Gond{\'a}n, L{\'a}szl{\'o} and Kocsis, Bence",
    title = "{Measurement Accuracy of Inspiraling Eccentric Neutron Star and Black Hole Binaries Using Gravitational Waves}",
    eprint = "1809.00672",
    archivePrefix = "arXiv",
    primaryClass = "astro-ph.HE",
    doi = "10.3847/1538-4357/aaf893",
    journal = "Astrophys. J.",
    volume = "871",
    number = "2",
    pages = "178",
    year = "2019"
}

@article{Zevin:2018kzq,
    author = "Zevin, Michael and Samsing, Johan and Rodriguez, Carl and Haster, Carl-Johan and Ramirez-Ruiz, Enrico",
    title = "{Eccentric Black Hole Mergers in Dense Star Clusters: The Role of Binary{\textendash}Binary Encounters}",
    eprint = "1810.00901",
    archivePrefix = "arXiv",
    primaryClass = "astro-ph.HE",
    reportNumber = "LIGO-P1800275",
    doi = "10.3847/1538-4357/aaf6ec",
    journal = "Astrophys. J.",
    volume = "871",
    number = "1",
    pages = "91",
    year = "2019"
}

@article{Yang:2024vfy,
    author = "Yang, Tao and Cai, Rong-Gen and Cao, Zhoujian and Lee, Hyung Mok",
    title = "{Enhancing early detection and localization of gravitational waves via eccentricity-induced higher harmonic modes with 2G detector networks}",
    eprint = "2412.20664",
    archivePrefix = "arXiv",
    primaryClass = "gr-qc",
    doi = "10.1007/s11433-025-2931-4",
    journal = "Sci. China Phys. Mech. Astron.",
    volume = "69",
    number = "6",
    pages = "260412",
    year = "2026"
}

@article{Yang:2022fgp,
    author = "Yang, Tao and Cai, Rong-Gen and Cao, Zhoujian and Lee, Hyung Mok",
    title = "{Parameter estimation of eccentric gravitational waves with a decihertz observatory and its cosmological implications}",
    eprint = "2212.11131",
    archivePrefix = "arXiv",
    primaryClass = "gr-qc",
    doi = "10.1103/PhysRevD.107.043539",
    journal = "Phys. Rev. D",
    volume = "107",
    number = "4",
    pages = "043539",
    year = "2023"
}

@article{Ma:2017bux,
    author = "Ma, Sizheng and Cao, Zhoujian and Lin, Chun-Yu and Pan, Hsing-Po and Yo, Hwei-Jang",
    title = "{Gravitational wave source localization for eccentric binary coalesce with a ground-based detector network}",
    eprint = "1710.02965",
    archivePrefix = "arXiv",
    primaryClass = "gr-qc",
    doi = "10.1103/PhysRevD.96.084046",
    journal = "Phys. Rev. D",
    volume = "96",
    number = "8",
    pages = "084046",
    year = "2017"
}

@article{Pan:2019anf,
    author = "Pan, Hsing-Po and Lin, Chun-Yu and Cao, Zhoujian and Yo, Hwei-Jang",
    title = "{Accuracy of source localization for eccentric inspiraling binary mergers using a ground-based detector network}",
    eprint = "1912.04455",
    archivePrefix = "arXiv",
    primaryClass = "gr-qc",
    doi = "10.1103/PhysRevD.100.124003",
    journal = "Phys. Rev. D",
    volume = "100",
    number = "12",
    pages = "124003",
    year = "2019"
}

@article{Sun:2015bva,
    author = "Sun, Baosan and Cao, Zhoujian and Wang, Yan and Yeh, Hsien-Chi",
    title = "{Parameter estimation of eccentric inspiraling compact binaries using an enhanced post circular model for ground-based detectors}",
    doi = "10.1103/PhysRevD.92.044034",
    journal = "Phys. Rev. D",
    volume = "92",
    number = "4",
    pages = "044034",
    year = "2015"
}

@article{Graham:2020gwr,
    author = "Graham, M. J. and others",
    title = "{Candidate Electromagnetic Counterpart to the Binary Black Hole Merger Gravitational Wave Event S190521g}",
    eprint = "2006.14122",
    archivePrefix = "arXiv",
    primaryClass = "astro-ph.HE",
    doi = "10.1103/PhysRevLett.124.251102",
    journal = "Phys. Rev. Lett.",
    volume = "124",
    number = "25",
    pages = "251102",
    year = "2020"
}

@article{DES:2019ccw,
    author = "Soares-Santos, M. and others",
    collaboration = "DES, LIGO Scientific, Virgo",
    title = "{First Measurement of the Hubble Constant from a Dark Standard Siren using the Dark Energy Survey Galaxies and the LIGO/Virgo Binary{\textendash}Black-hole Merger GW170814}",
    eprint = "1901.01540",
    archivePrefix = "arXiv",
    primaryClass = "astro-ph.CO",
    reportNumber = "FERMILAB-PUB-18-629-AE",
    doi = "10.3847/2041-8213/ab14f1",
    journal = "Astrophys. J. Lett.",
    volume = "876",
    number = "1",
    pages = "L7",
    year = "2019"
}

@article{LIGOScientific:2021aug,
    author = "Abbott, R. and others",
    collaboration = "LIGO Scientific, Virgo, KAGRA",
    title = "{Constraints on the Cosmic Expansion History from GWTC{\textendash}3}",
    eprint = "2111.03604",
    archivePrefix = "arXiv",
    primaryClass = "astro-ph.CO",
    reportNumber = "LIGO-P2100185-v6, LIGO-P2100185-v5",
    doi = "10.3847/1538-4357/ac74bb",
    journal = "Astrophys. J.",
    volume = "949",
    number = "2",
    pages = "76",
    year = "2023"
}

@misc{LIGOScientific:2025jau,
    author = "Abac, A. G. and others",
    collaboration = "LIGO Scientific, VIRGO, KAGRA",
    title = "{GWTC-4.0: Constraints on the Cosmic Expansion Rate and Modified Gravitational-wave Propagation}",
    eprint = "2509.04348",
    archivePrefix = "arXiv",
    primaryClass = "astro-ph.CO",
    reportNumber = "LIGO-P2400152",
    month = "9",
    year = "2025"
}

@article{Yang:2022tig,
    author = "Yang, Tao and Cai, Rong-Gen and Cao, Zhoujian and Lee, Hyung Mok",
    title = "{Eccentricity of Long Inspiraling Compact Binaries Sheds Light on Dark Sirens}",
    eprint = "2202.08608",
    archivePrefix = "arXiv",
    primaryClass = "gr-qc",
    doi = "10.1103/PhysRevLett.129.191102",
    journal = "Phys. Rev. Lett.",
    volume = "129",
    number = "19",
    pages = "191102",
    year = "2022"
}

@article{Yang:2022iwn,
    author = "Yang, Tao and Cai, Rong-Gen and Lee, Hyung Mok",
    title = "{Space-borne atom interferometric gravitational wave detections. Part III. Eccentricity on dark sirens}",
    eprint = "2208.10998",
    archivePrefix = "arXiv",
    primaryClass = "gr-qc",
    doi = "10.1088/1475-7516/2022/10/061",
    journal = "JCAP",
    volume = "10",
    pages = "061",
    year = "2022"
}

@article{Favata:2013rwa,
    author = "Favata, Marc",
    title = "{Systematic parameter errors in inspiraling neutron star binaries}",
    eprint = "1310.8288",
    archivePrefix = "arXiv",
    primaryClass = "gr-qc",
    reportNumber = "LIGO-P1300191",
    doi = "10.1103/PhysRevLett.112.101101",
    journal = "Phys. Rev. Lett.",
    volume = "112",
    pages = "101101",
    year = "2014"
}

@article{Narayan:2023vhm,
    author = "Narayan, Purnima and Johnson-McDaniel, Nathan K. and Gupta, Anuradha",
    title = "{Effect of ignoring eccentricity in testing general relativity with gravitational waves}",
    eprint = "2306.04068",
    archivePrefix = "arXiv",
    primaryClass = "gr-qc",
    doi = "10.1103/PhysRevD.108.064003",
    journal = "Phys. Rev. D",
    volume = "108",
    number = "6",
    pages = "064003",
    year = "2023"
}

@article{Saini:2022igm,
    author = "Saini, Pankaj and Favata, Marc and Arun, K. G.",
    title = "{Systematic bias on parametrized tests of general relativity due to neglect of orbital eccentricity}",
    eprint = "2203.04634",
    archivePrefix = "arXiv",
    primaryClass = "gr-qc",
    reportNumber = "LIGO Preprint No. P2200073",
    doi = "10.1103/PhysRevD.106.084031",
    journal = "Phys. Rev. D",
    volume = "106",
    number = "8",
    pages = "084031",
    year = "2022"
}

@article{KAGRA:2023pio,
    author = "Abbott, R. and others",
    collaboration = "KAGRA, VIRGO, LIGO Scientific",
    title = "{Open Data from the Third Observing Run of LIGO, Virgo, KAGRA, and GEO}",
    eprint = "2302.03676",
    archivePrefix = "arXiv",
    primaryClass = "gr-qc",
    reportNumber = "LIGO-P2200316",
    doi = "10.3847/1538-4365/acdc9f",
    journal = "Astrophys. J. Suppl.",
    volume = "267",
    number = "2",
    pages = "29",
    year = "2023"
}

@article{pesummary,
    author = "Hoy, Charlie and Raymond, Vivien",
    title = "{PESummary: the code agnostic Parameter Estimation Summary page builder}",
    eprint = "2006.06639",
    archivePrefix = "arXiv",
    primaryClass = "astro-ph.IM",
    reportNumber = "LIGO-P2000156",
    doi = "10.1016/j.softx.2021.100765",
    journal = "SoftwareX",
    volume = "15",
    pages = "100765",
    year = "2021"
}

@ARTICLE{2020SciPy-NMeth,
  author  = {Virtanen, Pauli and Gommers, Ralf and Oliphant, Travis E. and
            Haberland, Matt and Reddy, Tyler and Cournapeau, David and
            Burovski, Evgeni and Peterson, Pearu and Weckesser, Warren and
            Bright, Jonathan and {van der Walt}, St{\'e}fan J. and
            Brett, Matthew and Wilson, Joshua and Millman, K. Jarrod and
            Mayorov, Nikolay and Nelson, Andrew R. J. and Jones, Eric and
            Kern, Robert and Larson, Eric and Carey, C J and
            Polat, {\.I}lhan and Feng, Yu and Moore, Eric W. and
            {VanderPlas}, Jake and Laxalde, Denis and Perktold, Josef and
            Cimrman, Robert and Henriksen, Ian and Quintero, E. A. and
            Harris, Charles R. and Archibald, Anne M. and
            Ribeiro, Ant{\^o}nio H. and Pedregosa, Fabian and
            {van Mulbregt}, Paul and {SciPy 1.0 Contributors}},
  title   = {{{SciPy} 1.0: Fundamental Algorithms for Scientific
            Computing in Python}},
  journal = {Nature Methods},
  year    = {2020},
  volume  = {17},
  pages   = {261--272},
  adsurl  = {https://rdcu.be/b08Wh},
  doi     = {10.1038/s41592-019-0686-2},
}

@misc{GWOSC,
  howpublished = {\url{https://gwosc.org/}}
}

\end{document}